%% file: JoshiParkerWadsleyKeller2018.tex
\documentclass[a4paper,fleqn,usenatbib]{mnras}

\usepackage[T1]{fontenc}
\usepackage{ae,aecompl}

\usepackage{graphicx}	
\usepackage{amsmath}	
\usepackage{amssymb}	
\usepackage[svgnames]{xcolor}

\title[Mass loss and preprocessing of group galaxies]{The trajectories of galaxies in groups: mass loss and preprocessing}

\author[G.D. Joshi et al.]{Gandhali D. Joshi$^{1,2}$\thanks{Email: joshi@mpia-hd.mpg.de}, Laura C. Parker$^{2}$, James Wadsley$^{2}$, Benjamin W. Keller$^{2,3}$ \\
$^{1}$Max-Planck-Institute f{\"u}r Astronomie, 69117 Heidelberg, Germany \\
$^{2}$Department of Physics and Astronomy, McMaster University, Hamilton, ON L8S 4M1, Canada\\
$^{3}$Zentrum f{\"u}r Astronomie der Universit{\"a}t Heidelberg, ARI, 69120 Heidelberg, Germany}

\date{Accepted XXX. Received YYY; in original form ZZZ}

\pubyear{2018}

\begin{document}
\label{firstpage}
\pagerange{\pageref{firstpage}--\pageref{lastpage}}
\maketitle

\begin{abstract}
We present a study of environmental effects and preprocessing in a large galaxy group using a high-resolution, zoom-in simulation run with the \textsc{gasoline2} hydrodynamics code.  We categorize galaxies that were always in distinct haloes as \emph{unaccreted}, galaxies that were distinct before accretion onto the main group as \emph{single}, and galaxies that were in external sub-groups before accretion onto the main group as \emph{grouped}.

The unaccreted galaxy population experiences steady growth in dark matter, gas and stellar mass. Both single- and group-accreted galaxies begin to lose dark matter and gas after first accretion onto any host but continue to grow in stellar mass. Individual trajectories show that galaxies cease mass growth within roughly three virial radii of the main group. Single galaxies continue to form stars until the group virial radius is crossed, when they begin to lose both dark matter and gas. Grouped galaxies peak in mass when joining their external sub-group, indicating that they experience preprocessing. Most accreted galaxies retain their accumulated stellar mass. The total mass loss is dominated by tidal stripping, with evidence for additional gas stripping via ram pressure. Most accreted galaxies are quenched $\sim$(0.5-2.5) Gyr after accretion onto any group.

These differing histories place unaccreted, single and grouped galaxies in distinct regions of the stellar mass-to-halo mass (SMHM) relation. This suggests that preprocessed galaxies are a key source of scatter in the SMHM relation for mixed galaxy populations.
\end{abstract}




\section{Introduction}	\label{sec:intro}
The fate of galaxies as they evolve over time is strongly tied to their environment. Galaxies in dense environments, ranging from groups of a few galaxies to clusters of 100-1000s of galaxies, exhibit different properties when compared to isolated field galaxies. Group and cluster populations are dominated by red, elliptical and quenched galaxies, whereas field populations contain more blue, spiral, star forming galaxies \citep[see e.g.][]{Oemler74,Dressler80,Balogh04,Hogg04,Kauffmann04,Blanton05}. According to the hierarchical model for the growth of structure in the Universe, larger objects such as groups and clusters are built by the coalescence of smaller objects. As larger structures form, galaxies continue evolving in these environments. Understanding how these galaxies are transformed in larger structures is an important aspect of studying galaxy evolution.

There are several processes which can transform field galaxies as they approach and are accreted onto groups or clusters. Some processes are the result of a higher frequency of galaxy-galaxy interactions including mergers, which can cause starbursts that rapidly deplete the galaxy's fuel for future star formation (SF) \citep{Makino97,Angulo09}, and harassment, high speed encounters which can heat the galaxy's stars and dark matter, making them less bound and therefore easier to strip \citep{Moore96,Moore98}. Other effects are due to interactions with the group/cluster halo and halo gas such as tidal truncation, starvation and ram pressure stripping. Tidal truncation due to the group/cluster's stronger gravitation restricts the galaxy's own accretion and may also remove mass from the galaxy \citep{Toomre72,Barnes92,Bournaud04}. Starvation, the removal of the galaxy's hot gas halo, prevents future star formation on long timescales \citep{Larson80,Balogh00,Kawata08}. Finally, ram pressure stripping by the surrounding intragroup or intracluster medium (IGM or ICM) can remove even the more strongly bound cold gas from the galaxy, thereby rapidly stopping star formation \citep{Gunn72,Abadi99}. Each of these processes acts on different timescales and is efficient at different locations within the group/cluster halo. These processes also affect the various components of the galaxies differently and can lead to signature features in the proportions of dark matter, gas and stellar content in a galaxy. 

Galaxies are expected to lose mass in more massive haloes due to these environmental effects. \citet{Knebe06} used a suite of zoom-in dark matter simulations of galaxy clusters to show how tidal interactions between subhaloes can account for $\sim30\%$ of the mass loss experienced by cluster members. \citet{VanDenBosch16} carried out an extensive study of radial segregation in various properties of subhaloes in the Bolshoi and Chinchilla N-body simulations. They found a significant correlation between the $z=0$ distance of the subhaloes from the host centre and the amount of mass loss from the time of accretion. \citet{Behroozi14} also examined the trajectories of galaxy haloes in the Bolshoi simulation and found that such haloes can begin losing mass well outside the final host halo, at a median distance of $\sim2\,r_{\text{vir}}$, and going as far out as $\sim4\,r_{\text{vir}}$.

Larger clusters are gradually built up through the accretion of smaller groups of galaxies. These galaxies may have already been affected by their group environment, i.e. they may have been preprocessed, before they become a part of the final cluster. For example, \citet{McGee09} found that a large fraction of galaxies in cluster simulations were accreted as part of smaller groups and were therefore potentially preprocessed. They found that more massive galaxies and galaxies previously residing in more massive haloes have a higher probability of being preprocessed. In the lower mass regime, \citet{Wetzel15} studied galaxies in analogues of the Local Group, via the \textsc{elvis} simulations and semi-analytical models. They found that $\sim25\%$ of all satellites at $z=0$ had been preprocessed in groups of masses $M_{\text{vir}}>10^{11}M_{\sun}$.

Separating the effects of the final cluster from those of the preprocessing groups is crucial in determining the efficiency of the various environmental effects. \citet{Bahe13} examined several galaxy properties and radial trends in properties, in the \textsc{gimic} suite of simulations. When examining galaxies in massive clusters, they found strong radial trends in the proportion of galaxies that had sSFRs, hot gas fractions, and cold gas fractions above a certain threshold. However, they also found that $\sim50\%$ of those galaxies had been accreted as part of smaller groups; when these were excluded, the radial trends were significantly weaker.  \citet{Hou14} studied groups and clusters using SDSS data and found that galaxies infalling in substructure had higher quenched fractions. \citet{Gabor15} used simulations to study galaxy clusters and found that roughly 1/3 of galaxies had already been quenched in groups of mass $>10^{12}M_{\sun}$.

In our previous work \citep{Joshi17}, we investigated the halo mass loss of galaxy analogues in a dark matter simulation. We found that galaxies that had been in a group prior to accretion lost $\sim30\%$ more mass, relative to their peak mass, compared to galaxies that were distinct before accretion. The mass loss was found to strongly correlate with the time spent in a dense environment, be it the final host halo or any previous group halo in which the galaxies may have been preprocessed. 

In this work, we extend this previous work by producing a high-resolution, hydrodynamical simulation that incorporates key baryonic processes. We simulate a single group of galaxies and study in detail the accretion histories of individual galaxies. We examine the mass loss experienced by each galaxy as a result of being part of a group, the impact on star formation, and whether there is evidence for preprocessing of these galaxies. The details of the simulation and selection of galaxies are provided in Section \ref{sec:methods}. In Section \ref{sec:evol}, we examine the average trends in mass loss over cosmic time for galaxies that were accreted onto the main group as single galaxies vs. those that had previously been part of a group. In Section \ref{sec:traj}, we follow the galaxies individually as they approach the group and examine their dark matter, gas and stellar content. We discuss what may be the dominant processes driving the mass evolution of these galaxies in Section \ref{sec:lossMech}. Finally, we explore the consequences of such evolution on the final properties of our galaxy samples in Section \ref{sec:implications}. Our findings are summarized in Section \ref{sec:summ}.

\section{Methods}	\label{sec:methods}

\subsection{Simulation}
We carry out a high-resolution hydrodynamical zoom-in simulation of a galaxy group using \textsc{gasoline2} \citep{Wadsley17}. \textsc{gasoline2} is a smoothed particle hydrodynamics (SPH) code that includes prescriptions for star formation, radiative and metal line gas cooling, supernovae and stellar winds. It also employs the superbubble feedback model of \citet{Keller14}. We first ran a lower resolution cosmological N-body simulation run using \textsc{changa} \citep{Jetley08,Jetley10,Menon14} in gravity-only mode. This first simulation was comprised of a (100 Mpc)$^3$ comoving volume containing $1024^3$ particles, resulting in a particle mass of $3.7\times10^7 M_{\sun}$. The initial conditions (ICs) were generated at $z=100$ using the code \textsc{music} \citep{Hahn13}, assuming a flat standard $\Lambda$CDM cosmology with $\Omega_{\Lambda} = 0.6914$, $\Omega_{\text{m}} = 0.3086$, $h = 0.6777$, $n_{\text{s}} = 0.9611$ and $\sigma_{8} = 0.8288$ \citep{Planck14}. The simulation was run over 1000 timesteps, linear in time, to $z=0$. We identified haloes in the simulation at $z=0$ using the phase-space friends-of-friends (FOF) algorithm \textsc{rockstar} \citep{Behroozi13a}. One of these final haloes was then chosen for the zoom-in simulation.

To focus on the galaxy group regime, we restricted the sample to distinct isolated haloes with masses within $(2-3)\times10^{13}M_{\sun}$. This mass selection provided a few tens of groups. The final group was selected at random from the resulting set of groups, although we visually confirmed that the group had at least a few subhaloes that were distinct from the group halo and could host galaxies. The group had a virial mass of $2.7\times10^{13} \text{M}_{\sun}$ and a virial radius of 789 kpc at $z=0$ in the dark matter run.

In order to generate ICs for the zoom-in simulation, all particles within $3\,r_{\text{vir}}$ from the group centre at $z=0$ were selected and tracked back to their positions in the ICs of the original dark matter simulation. Cosmological zoom-in simulations work by identifying the high-redshift Lagrangian progenitor volume of a halo identified at late times. The region that must be refined often has a complex geometry, making the choice of refining volumes non-trivial. Simulating high-resolution material that ends up outside the halo is expensive both in terms of computing time and storage requirements. We use a new, efficient algorithm for generating zoom-in initial conditions. We start with a grid at the lowest refinement/resolution level. Next, grid cells which contain any particles identified as part of the zoom region are set to the highest resolution. Particles in a shell around this high-resolution region are set to the next highest resolution, a factor of 2 lower than the previous. This process is repeated until we reach the lowest resolution of the IC. This gives us a ``voxelized'' refinement region, with an optimal high-resolution volume surrounded by increasingly lower-resolution shells. This algorithm is similar to the strategy of generating nested grids in Adaptive Mesh Refinement (AMR) simulations \citep{Berger84}. We implement this algorithm in \textsc{music} and find that it produces ICs with smaller high resolution regions compared to a more typical convex hull configuration (a convex closed surface containing the particles of interest), with the resulting high-resolution region containing 2-3 times fewer particles than a convex hull configuration would have produced. In the ICs we generated for the zoom-in simulation, we refine from an \emph{effective} resolution (for a $(100\  \text{Mpc})^{3}$ volume) of $256^{3}$ particles in the lowest resolution regions to $2\times2048^{3}$ particles in the highest resolution regions (note that baryonic particles are only generated in the high-resolution region). Within the lowest resolution region, the dark-matter particle mass is $2.4\times10^9 \text{M}_{\sun}$, whereas within the high-resolution region, the dark-matter particle mass is $3.9\times10^6 \text{M}_{\sun}$ and the baryon particle mass is $7.2\times10^5 \text{M}_{\sun}$.

These initial conditions were then used to run the zoom-in hydrodynamical simulation from redshift $z=100$ to $z=0$ in 1024 timesteps, with every eighth snapshot saved. This gives us 128 snapshots equally spaced in time, with consecutive snapshots separated by 107.64 Myr. At $z=0$, the group has a virial mass of $1.87\times10^{13} \text{M}_{\sun}$ and a virial radius of 665 kpc in the hydrodynamical run. Its mass in dark matter, gas and stars is $1.57\times10^{13} \text{M}_{\sun}$, $1.80\times10^{12} \text{M}_{\sun}$ and $1.17\times10^{12} \text{M}_{\sun}$ respectively.

\subsection{Halo finding and galaxy properties}

\begin{figure}
\includegraphics[width=\linewidth]{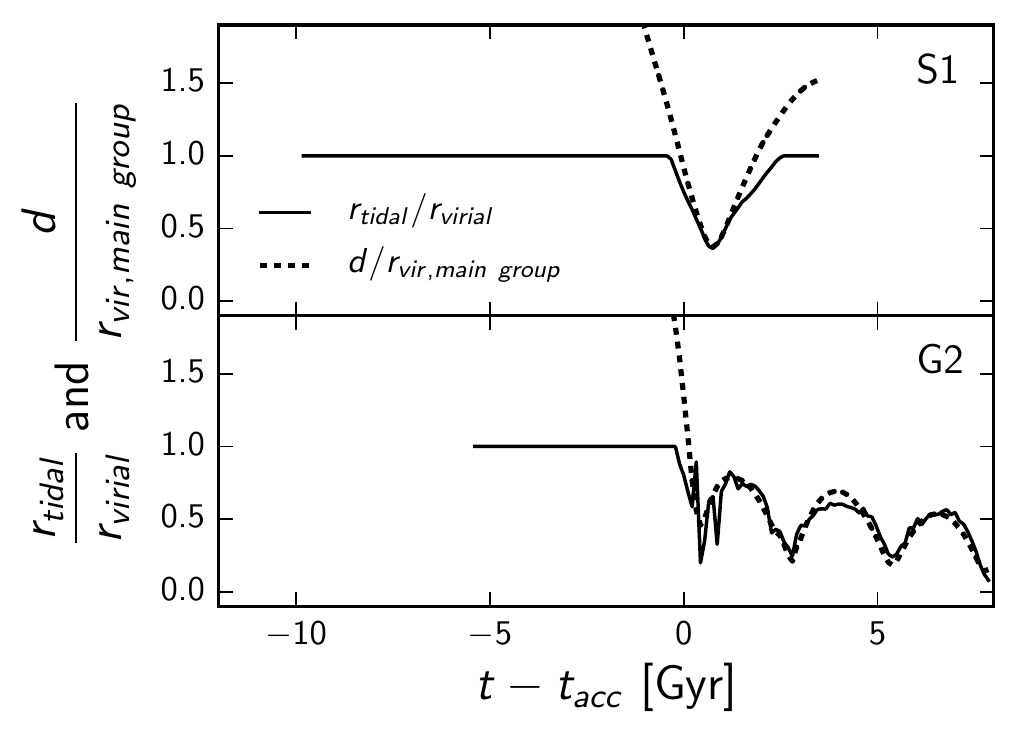}
\caption{The ratio of tidal radius to virial radius (solid line) for two typical galaxies, and distance to the main group (dotted line), as a function of time since first accretion. The distance is normalized by the main group's instantaneous virial radius.}\label{fig:rtideDistVsTime}
\end{figure}

\begin{figure*}
\includegraphics[width=\linewidth]{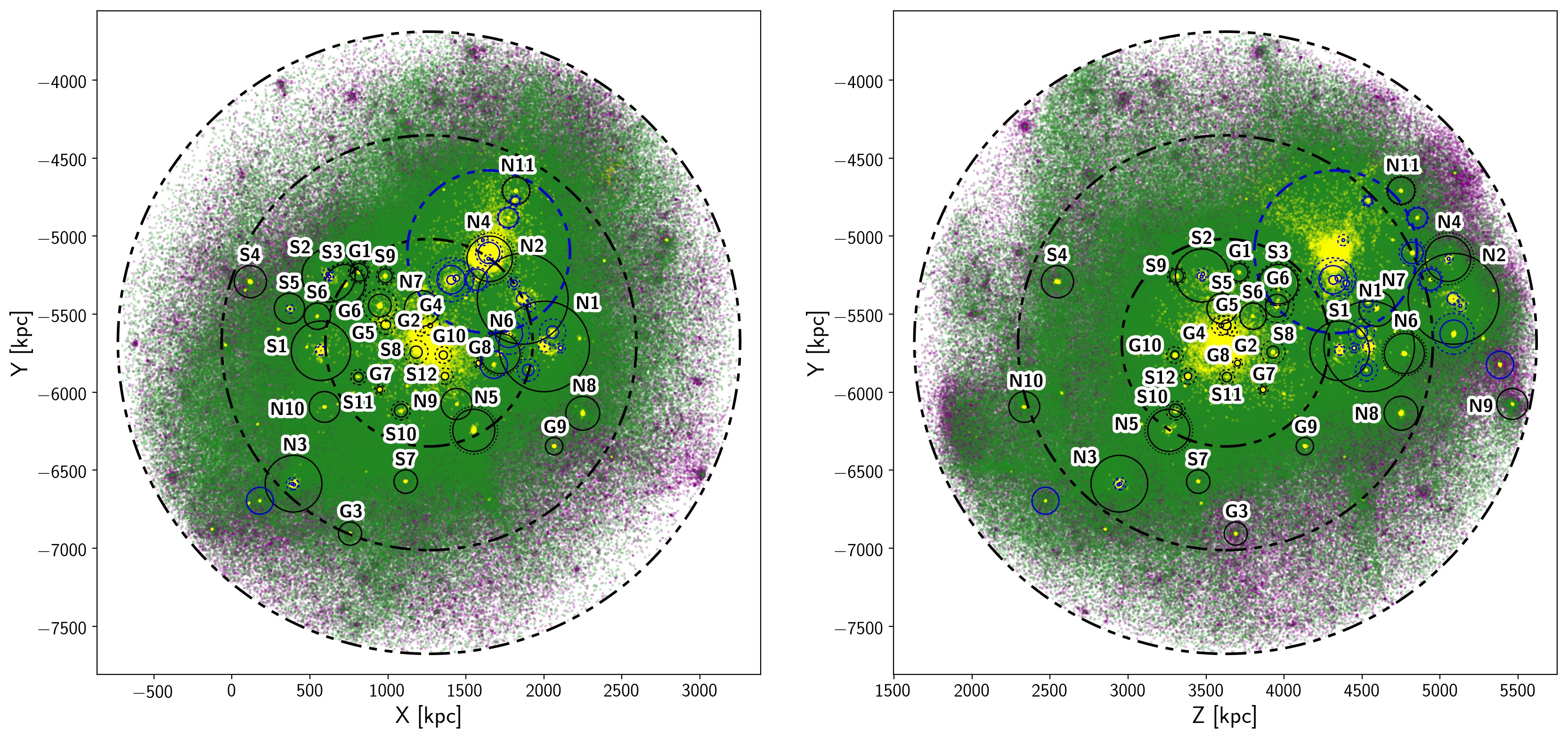}
\caption{Two projections of the group at $z=0$ showing its mass distribution and the positions of our galaxy samples. The coloured points are particles within $3\,r_{\text{vir}}$ of the group; dark matter particles are shown in purple, gas particles in green and star particles in yellow (we only plot every tenth particle for clarity). The black dash-dot lines show $r_{\text{vir}}$, $2\,r_{\text{vir}}$ and $3\,r_{\text{vir}}$ of the main group. The tidal radius of each galaxy is shown by the black solid circles and their virial radius is shown by the black dotted circles. Note that in practice, only galaxies within $\sim1.3\ \times$ a neighbouring galaxy's virial radius are truncated; for all other galaxies, the tidal and virial radii are equal, and hence the dotted and solid circles lie on top of each other. Galaxy labels beginning with `N' denote unaccreted galaxies, those beginning with `S' denote singly accreted galaxies and those beginning with `G' denote group accreted galaxies. The second external group as well as any galaxies that were part of other groups but never accreted by the main group are shown by the blue circles. These galaxies are excluded from our analysis.}	\label{fig:galPos}
\end{figure*}

\begin{table*}
\caption{Galaxy properties for the never accreted, singly-accreted and group-accreted samples. The reported radii and masses [cols 2-6] are at $z=0$. $d_{\text{min}}$ is the closest distance to the main group that the galaxy reaches (normalized by the group's virial radius at the time, $r_{\text{vir}}$). $z_{\text{acc}}$ is the redshift at which the single and grouped galaxies first cross within the tidal radius of the main group and $z_{\text{group}}$ is the redshift at which the grouped galaxies first become part of an external group (that is not part of the main group's subhalo hierarchy).}	\label{tab:finalProps}
\begin{tabular}{lcccccccc}
\hline
\multicolumn{1}{l}{Label} & \multicolumn{1}{c}{$r_{\text{tid}}$} & \multicolumn{1}{c}{$M_{\text{tot}}$} & \multicolumn{1}{c}{$M_{\text{dm}}$} & \multicolumn{1}{c}{$M_{\text{gas}}$} & \multicolumn{1}{c}{$M_{\text{star}}$} & \multicolumn{1}{c}{$d_{\text{min}}$} & \multicolumn{1}{c}{$z_{\text{acc}}$} & \multicolumn{1}{c}{$z_{\text{group}}$}\\
& \multicolumn{1}{c}{[kpc]} & \multicolumn{1}{c}{[$M_{\sun}$]} & \multicolumn{1}{c}{[$M_{\sun}$]} & \multicolumn{1}{c}{[$M_{\sun}$]} & \multicolumn{1}{c}{[$M_{\sun}$]} & \multicolumn{1}{c}{[$r_{\text{vir}}$]} & & \\
\hline
\multicolumn{9}{c}{Never accreted}	\\
\hline
\input{figures/distinct_properties}
\hline
\multicolumn{9}{c}{Single accretion}	\\
\hline
\input{figures/single_properties}
\hline
\multicolumn{9}{c}{Grouped accretion}	\\
\hline
\input{figures/grouped_properties}
\hline
\end{tabular}
\end{table*}

\begin{table*}
\caption{Galaxy properties at $z_{\text{peak}}$, the redshift at which the galaxy's \emph{total} mass is at its peak value. $d_{\text{peak}}$ is the galaxy's distance from the main group at $z_{\text{peak}}$ (normalized by the group's virial radius $r_{\text{vir}}$ at the time).}	\label{tab:peakProps}
\begin{tabular}{lccccccc}
\hline
\multicolumn{1}{l}{Label} & \multicolumn{1}{c}{$z_{\text{peak}}$} & \multicolumn{1}{c}{$d_{\text{peak}}$} & \multicolumn{1}{c}{$r_{\text{tid}}$} & \multicolumn{1}{c}{$M_{\text{tot}}$} & \multicolumn{1}{c}{$M_{\text{dm}}$} & \multicolumn{1}{c}{$M_{\text{gas}}$} & \multicolumn{1}{c}{$M_{\text{star}}$} \\
& & \multicolumn{1}{c}{[$r_{\text{vir}}$]} & \multicolumn{1}{c}{[kpc]} & \multicolumn{1}{c}{[$M_{\sun}$]} & \multicolumn{1}{c}{[$M_{\sun}$]} & \multicolumn{1}{c}{[$M_{\sun}$]} & \multicolumn{1}{c}{[$M_{\sun}$]} \\
\hline
\multicolumn{8}{c}{Never accreted}	\\
\hline
\input{figures/distinct_properties_peak}
\hline
\multicolumn{8}{c}{Single accretion}	\\
\hline
\input{figures/single_properties_peak}
\hline
\multicolumn{8}{c}{Grouped accretion}	\\
\hline
\input{figures/grouped_properties_peak}
\hline
\end{tabular}
\end{table*}

Haloes are identified using \textsc{rockstar} on the dark-matter particles. We then use \textsc{consistent trees} \citep{Behroozi13b} to generate robust merger trees for the haloes. \textsc{rockstar} has been shown to be reliable at tracking haloes even in dense environments due to its use of both particle positions and velocities \citep[e.g.][]{Knebe11,Behroozi13a,Joshi16}. The resulting halo catalogue provides several spherical overdensity properties for the haloes. The halo virial radius, $r_{\text{vir}}$, is defined as the radius within which the average density is $\Delta_{\text{c}}$ times the critical density of the Universe. The virial overdensity factor $\Delta_{\text{c}}$ is defined by \cite{Bryan98}:
\begin{equation}
\Delta_{\text{c}} = 18\pi^{2}+82x-39x^{2}
\end{equation}
where
\begin{equation}
x=\frac{\Omega_{\text{m,0}}(1+z)^{3}}{\Omega_{\text{m,0}}(1+z)^{3}+\Omega_{\Lambda}}-1
\end{equation}
For the cosmological parameters used in this study, $\Delta_\text{c}=102$ at $z=0$.

Following the halo finding process, we retain haloes with $M_{\text{vir}}>10^{9}M_{\sun}$ at $z=0$ (i.e. resolved by at least $250$ dark matter particles) and $M_{\text{vir}}>10^{8}M_{\sun}$ at higher redshifts to track progenitors.  We must then assign baryonic particles to these haloes.  \textsc{rockstar} provides a mass associated with each halo, which it labels the virial mass, $M_{\text{vir}}$.   We can derive a virial radius from this mass but to account for the influence of massive neighbours and the fact that some of the halos are subhalos of larger systems, we also determine a tidal radius for each halo. We calculate all other properties within this tidal radius.

We calculate the tidal radius as follows:
\begin{align} \label{eq:tidalR}
\frac{G\,m_{\text{halo}}(<r_{\text{tidal}})}{r_{\text{tidal}}^2} &= \frac{G\,M_{\text{neighbour}}\left(<(d-r_{\text{tidal}})\right)}{(d-r_{\text{tidal}})^2}\nonumber\\ &\qquad - \frac{G\,M_{\text{neighbour}}(<d)}{d^2}
\end{align}
where $m_{\text{halo}}$ and $M_{\text{neighbour}}$ are the masses of the halo itself and a neighbouring halo that can cause tidal truncation, and $d$ is the distance between them. We consider any distinct halo more massive than the halo in question and where $d<2\,r_{\text{vir,neighbour}}$ as a neighbour. If the halo in question is a subhalo, we also consider its direct host halo and the most massive host halo it belongs to (if different), as provided by \textsc{rockstar}. If there is more than one neighbour that could cause tidal truncation, we calculate several tidal radii corresponding to each neighbour and then take the smallest of these to be the halo's tidal radius.  

In figure~\ref{fig:rtideDistVsTime} we show the behaviour of the tidal radius over time for two typical accreting galaxies (S1 and G2; these labels are explained later in this section) compared to their distance from the main halo. The tidal radius is approximately the same as the virial radius just prior to accretion and rapidly shrinks in a manner that tracks the distance scaled to the virial radius of the main halo.  This behaviour can be understood via a simple linear model where the mass of a halo $M(<r) = M_{\text{vir}} \times (r/r_{\text{vir}})$ (which would be correct for a truncated isothermal sphere whose density scales as $\rho\sim r^{-2}$). When substituted into equation~\ref{eq:tidalR}, we recover a linear dependence: $r_{\text{tidal}} =  r_{\text{vir,halo}} \times (d/r_{\text{vir,neighbour}})$. The behaviour of actual accreting galaxies, whose density profiles may not be identical to an isothermal sphere, is quite similar. In the case of S1 we see a backsplash event where outer material is temporarily unbound then reassociated with the galaxy. For G2, the galaxy progressively losses orbital energy via dynamical friction. In both cases the tidal radius tracks the current group-centric distance.

Using a tidal radius is more physically motivated by the environment we are studying than the \textsc{rockstar} virial radius definition. The tidal radius fluctuates after accretion and gets especially small during pericentric passages, as shown in figure~\ref{fig:rtideDistVsTime}, which causes the bound mass to fluctuate rapidly but smoothly. As seen in the following sections, this leads to fluctuations in the bound mass versus time but key observables such as stellar mass are mostly unaffected.

Next we determine which particles are bound to the subhalo, using an approximate averaged spherical potential. The approach follows eq. 2.28 from \citep{Binney08} with three modifications: (i) singularities in the potential near the centre are avoided via a gravitational softening (length $\epsilon$), (ii) the second, integral term is only calculated up to the tidal radius, and (iii) we add an additional constant of integration to ensure that particles at the tidal radius with zero velocity are marginally unbound. Thus, particles are considered bound if they satisfy,
\begin{align}	\label{eq:spEnergy}
\frac{1}{2}v^2 &- \frac{Gm_\text{{halo}}(<r)}{\max(r,\epsilon)}\nonumber\\
&-4\pi G\int_{r}^{r_{\text{tidal}}}\frac{\rho_{\text{halo}}(r')r'^{2}dr'}{\max(r',\epsilon)} + \frac{Gm_{\text{tidal}}}{r_{\text{tidal}}}< 0
\end{align}
where $v$ is the velocity of the particles relative to the halo's bulk velocity and $\epsilon = 0.625$ kpc. In cases where subhaloes at the same level in the halo hierarchy have overlapping volumes, the particles are assigned to the subhalo to which they are most bound. We use these bound particles to measure all properties such as total mass and mass in dark matter, gas and stars. Only haloes with a total bound mass of $M_{\text{tot}}>10^{8}M_{\text{\sun}}$ or a bound stellar mass of $M_{*}>10^{7}M_{\text{\sun}}$ are retained.

Once the galaxy properties are determined, we generate histories for each galaxy that is within $3\,r_{\text{vir}}$ of the main group at $z=0$. Galaxy histories are determined by following the most massive progenitor in the merger trees. Our final sample of galaxies includes any galaxy which had a stellar mass of $M_{*}>10^{8}M_{\text{\sun}}$ during at least one previous snapshot and could be tracked back to at least $z=2$. This gives us a total of 52 galaxies, two of which are the main group central and the central of a nearby group. 17 other galaxies were never accreted by the main group, but were part of a different group at some point in their history. These galaxies, as well as the two group centrals, were not the focus of this study and are not included in further analysis.

Since we are interested in the effects of preprocessing, we separate the galaxies into three categories.
\begin{enumerate}
\item \textbf{Never accreted}: Galaxies that have been distinct for their entire history (hereafter referred to as `unaccreted' galaxies). 11 galaxies fall into this category.
\item \textbf{Single accretion}: Galaxies that were distinct for their entire history before they were accreted by the main group (hereafter referred to as `single' galaxies). 12 galaxies fall into this category.
\item \textbf{Grouped accretion}: Galaxies that were part of a different group for some time before they were accreted by the main group (hereafter referred to as `grouped' galaxies). 10 galaxies fall into this category.
\end{enumerate}
Fig. \ref{fig:galPos} shows the distribution of particles in the group as well as the galaxies in our sample. Table \ref{tab:finalProps} provides key details regarding our galaxy sample.

\section{Evolution of galaxy properties}	\label{sec:evol}

\begin{figure}
\includegraphics[width=\linewidth]{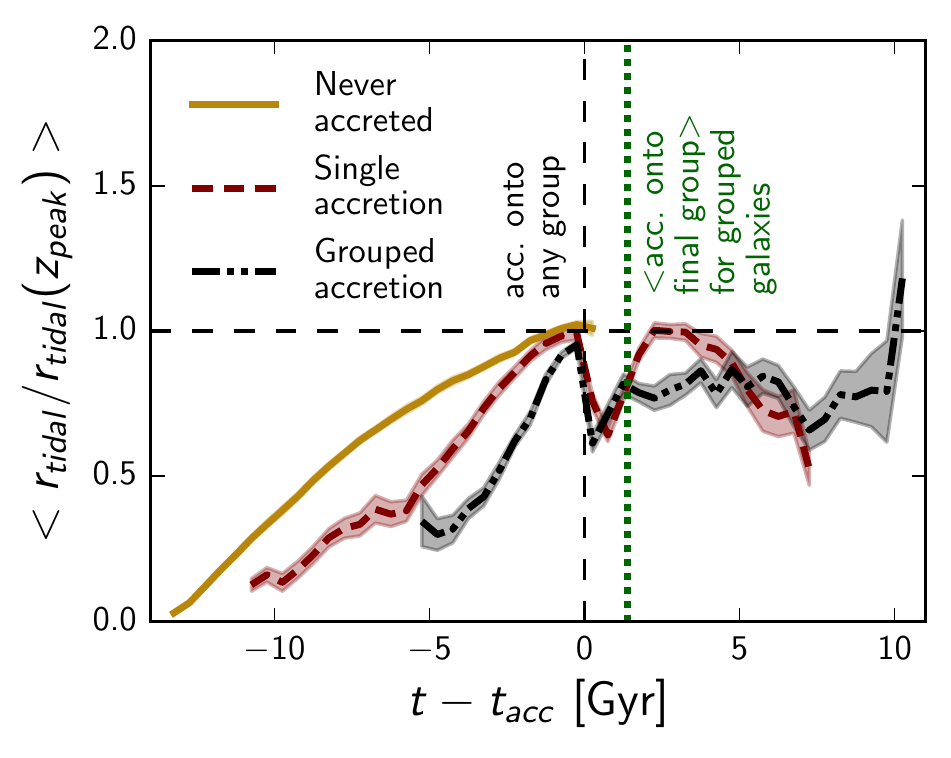}
\caption{Average galaxy tidal radius normalized by radius at peak total mass, as a function of time since first accretion (onto any group) in bins of 0.5 Gyr, for each galaxy sample. The coloured shaded regions show the standard uncertainty in the mean (note that the spread of the data is larger). The green dotted line indicates the average time at which the grouped galaxies first entered the main group.}	\label{fig:rtideEvol}
\end{figure}

\begin{figure}
\includegraphics[width=\linewidth]{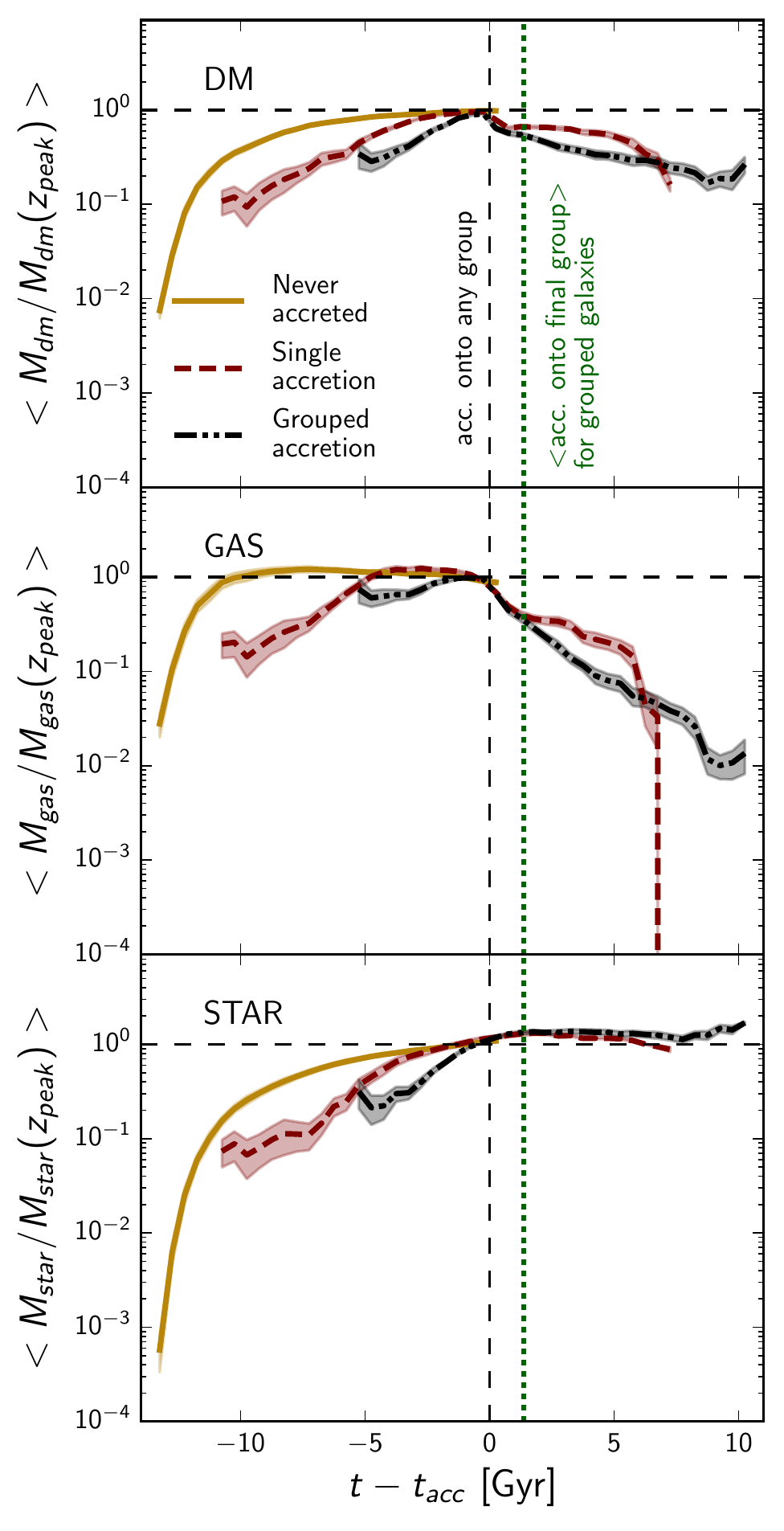}
\caption{Average mass in dark matter (top), gas (middle) and stars (bottom), normalized by their respective values at peak total mass, as a function of time since first accretion for each galaxy sample. The green dotted line indicates the average time at which the grouped galaxies first entered the main group as in Fig. \ref{fig:rtideEvol}. While the single and grouped galaxies lose mass in gas and dark matter after reaching a peak value, their stellar mass is largely unaffected.}	\label{fig:massEvolNorm}
\end{figure}

\begin{figure}
\includegraphics[width=\linewidth]{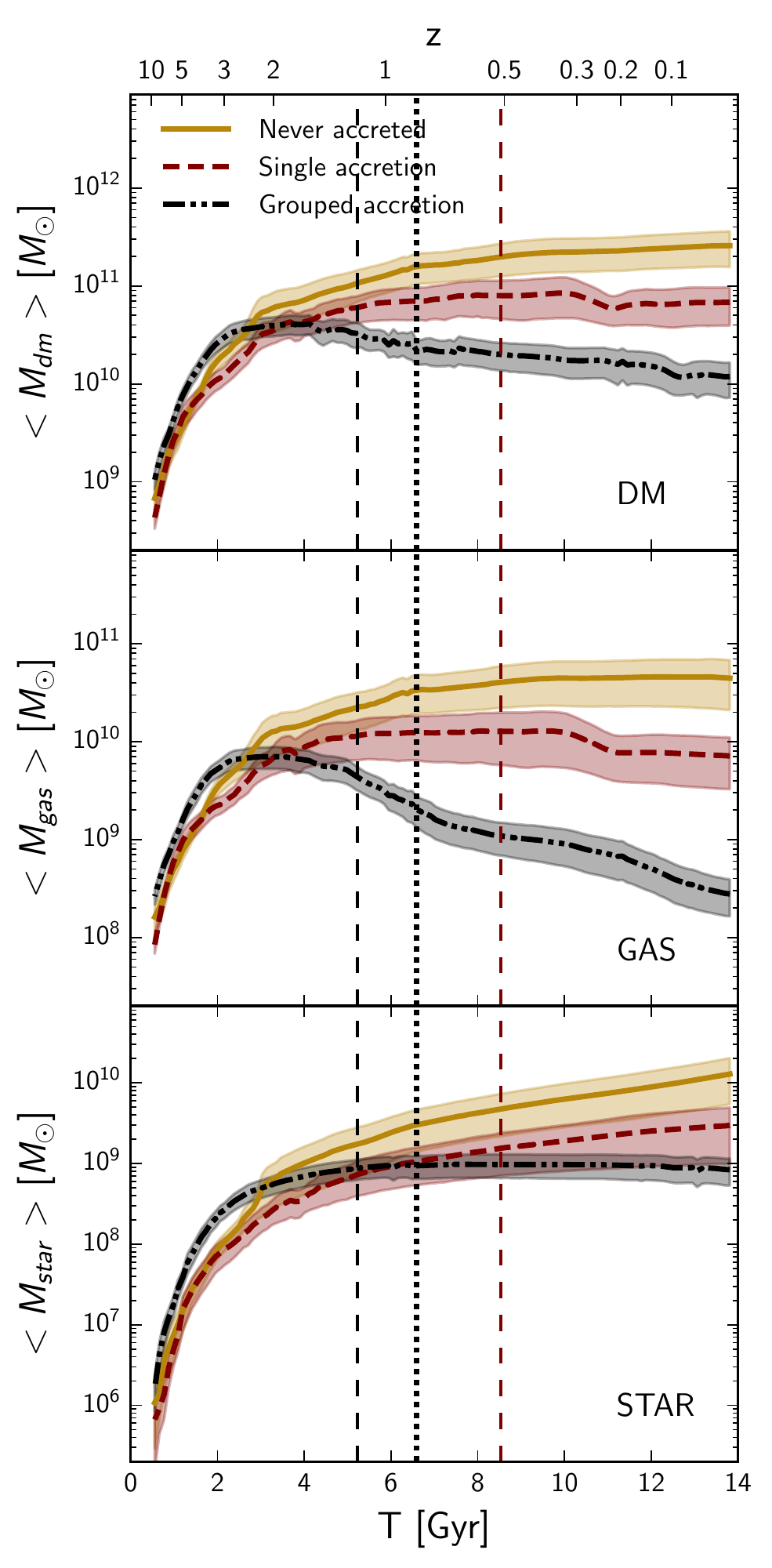}
\caption{Average mass in dark matter (top), gas (middle) and stars (bottom), as a function of cosmic time for each galaxy sample. The dashed red and black lines indicate the average time at which the single and grouped galaxies respectively are first accreted onto any group. The dotted black line indicates when the grouped galaxies are accreted onto the main group on average.}	\label{fig:massEvol}
\end{figure}

\begin{figure}
\includegraphics[width=\linewidth]{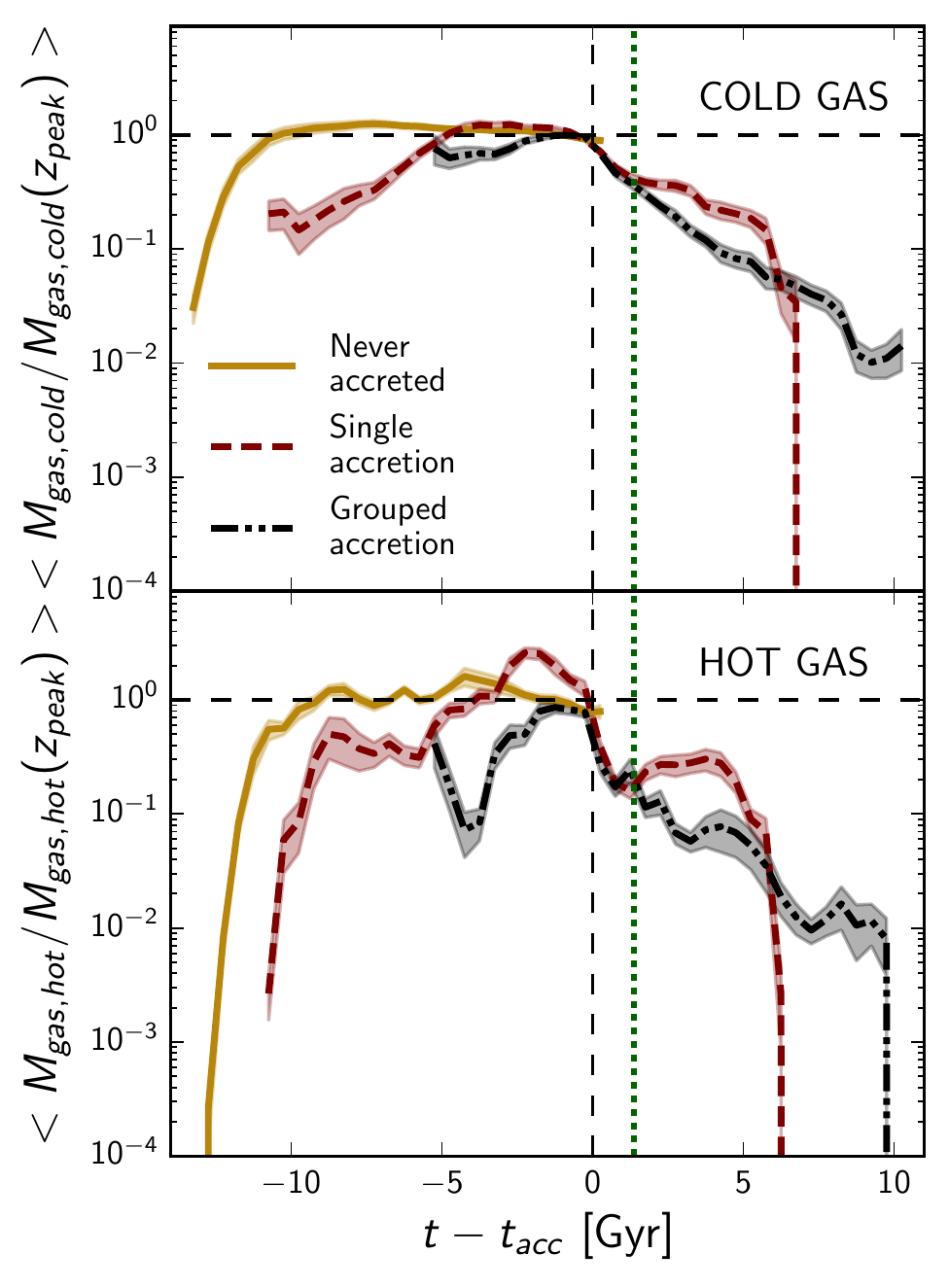}
\caption{Average mass in cold gas (top) and hot gas (bottom), normalized by their respective values at peak total mass, as a function of time since first accretion for each galaxy sample. Hot (cold) gas is defined here as having $T>$ ($<$) $10^5$K. The green dotted line indicates the average time at which the grouped galaxies first entered the main group as in Fig. \ref{fig:rtideEvol}.}	\label{fig:gasEvol}
\end{figure}

\begin{figure}
\includegraphics[width=\linewidth]{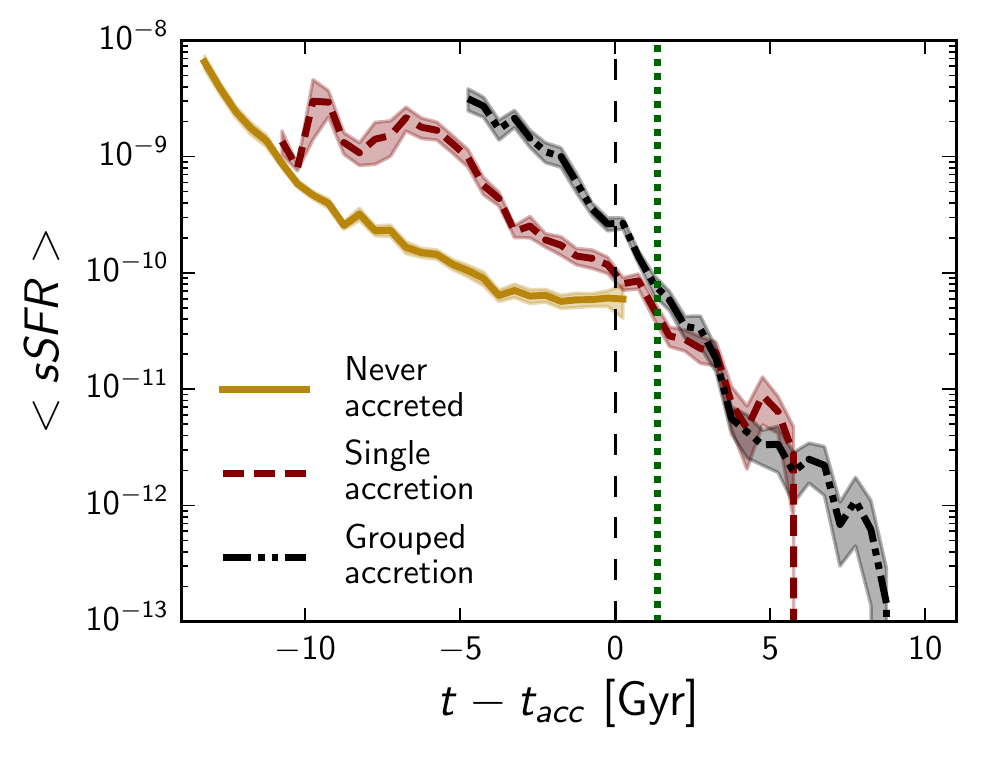}
\caption{Average sSFR as a function of time since first accretion for each galaxy sample. The green dotted line indicates the average time at which the grouped galaxies first entered the main group as in Fig. \ref{fig:rtideEvol}.}	\label{fig:ssfrEvol}
\end{figure}

We first examine the average evolution of properties for the unaccreted, singly accreted, and group accreted galaxy samples. As shown in Table \ref{tab:finalProps}, the three samples of galaxies have a wide range of sizes and masses. The single and grouped galaxies also have a large range of accretion redshifts ($z_{\text{acc}}$). Several studies have shown that a key parameter in determining the impact of environmental processes is the time since accretion \citep[e.g. see][]{Wetzel13,Joshi17,Rhee17,Armitage18}. We therefore examine the trends in various galaxy properties as a function of $t-t_{\text{acc}}$, where $t$ is the age of the Universe and $t_{\text{acc}}$ is the time at which the single and grouped galaxies were accreted onto \emph{any} group. For the unaccreted galaxies, we set $t_{\text{acc}}=t_{0}$, the current age of the Universe. Due to the large range in masses, we normalize the galaxy properties by their value at $z_{\text{peak}}$, the redshift at which the galaxy's \emph{total} mass is at its highest value. Table \ref{tab:peakProps} provides key properties of the galaxies at $z_{\text{peak}}$. We show the average results for all times where we have data for more than one galaxy. Fig. \ref{fig:rtideEvol} shows the average tidal radius of galaxies in each sample as a function of time since accretion. According to our definition, unaccreted galaxies are never truncated; single and grouped galaxies, however, show significant tidal truncation at infall.

In Fig. \ref{fig:massEvolNorm}, we show the evolution of the galaxies' mass in dark matter, gas and stars. Unaccreted galaxies continue to grow in mass in dark matter and stars at all times, while their average gas mass remains approximately constant after reaching a peak value. By comparison, the single and grouped galaxies both show a steady decline in dark matter and gas since just before accretion onto any group halo. This mass loss is due to tidal stripping, which is the one process by which galaxies can lose dark matter, and also since we are, by definition, examining the mass within the tidal radius here. The same trend is not seen in stellar mass; both the single galaxies and grouped galaxies have nearly constant stellar masses after accretion. The trends seen in Fig. \ref{fig:massEvol} indicate that while tidal stripping is efficient at removing the dark matter and gas content of galaxies, it does not affect the stellar content to the same degree. This is to be expected since tidal stripping is a process that works outside-in, first affecting the outskirts of the galaxy, and gradually moving inwards. It will therefore remove the more extended dark matter and gas components before reaching the more compact and tightly bound stellar component. In Fig. \ref{fig:massEvol}, we show the average evolution of the galaxies without any normalization in time or mass to examine the global trends present in our samples. It can be seen more clearly here that grouped galaxies begin losing dark matter before they are accreted onto any group. The single galaxies have nearly constant dark matter and gas masses, and show only modest stripping after accretion.

We next look at the details of the galaxies' gas content and star formation. Fig. \ref{fig:gasEvol} shows the average mass in cold and hot gas, where we use $10^{5}$K as the temperature at which we separate the two components. Note that the `cold' gas includes the warm ISM component of a galaxy along with the truly cold neutral medium. The cold gas is the larger gas component by mass, accounting for nearly all of the gas content of the galaxies in our sample, and hence its evolution effectively mirrors the evolution of the total gas mass. The results in Fig. \ref{fig:gasEvol} show that the galaxies are gradually being depleted of their fuel for star formation on fairly long timescales. The rate of decline in both gas components is similar to the loss of total gas mass for both single and grouped galaxies.

The star formation behaviour of the three galaxy samples is shown in Fig. \ref{fig:ssfrEvol}. Note that we do not normalize the sSFR by its value at peak mass. Before accretion, all three populations begin with the same sSFR of $\sim5\times10^{-9}\,yr^{-1}$ and show a steady decline in sSFR, roughly linear with time in the case of the single and grouped galaxies. After accretion, both the single and grouped galaxies show a continued decline in sSFR, with their rates of decline remaining roughly constant.

As mentioned above, since we explicitly use tidal radii to determine the masses of our galaxies, the mass loss we see, especially in the case of dark matter, is predominantly due to tidal stripping. Any additional mass loss can partly be explained as the consumption of gas by star formation, while the rest may be due to other processes such as ram pressure stripping, which we explore in Section \ref{sec:lossMech}. Figs. \ref{fig:massEvol}, \ref{fig:gasEvol} and \ref{fig:ssfrEvol} indicate that tidal stripping causes significant mass loss for single and grouped galaxies due to their dense environments and while this tidal stripping does not cause significant stellar mass loss, the reduced overall mass does appear to affect the grouped galaxies' sSFRs.

It is clear from Figs. \ref{fig:massEvol}, \ref{fig:gasEvol} and \ref{fig:ssfrEvol}, that in addition to the environmental effects that drive the differences between the average trend of unaccreted and singly accreted galaxies, preprocessing plays a significant role in further altering the properties of the grouped galaxies. In the next section, we explore the histories of each of the single and grouped galaxies individually to explicitly study the impact of preprocessing.

\section{Radial trajectories}	\label{sec:traj}

\begin{figure*}
\includegraphics[width=\linewidth]{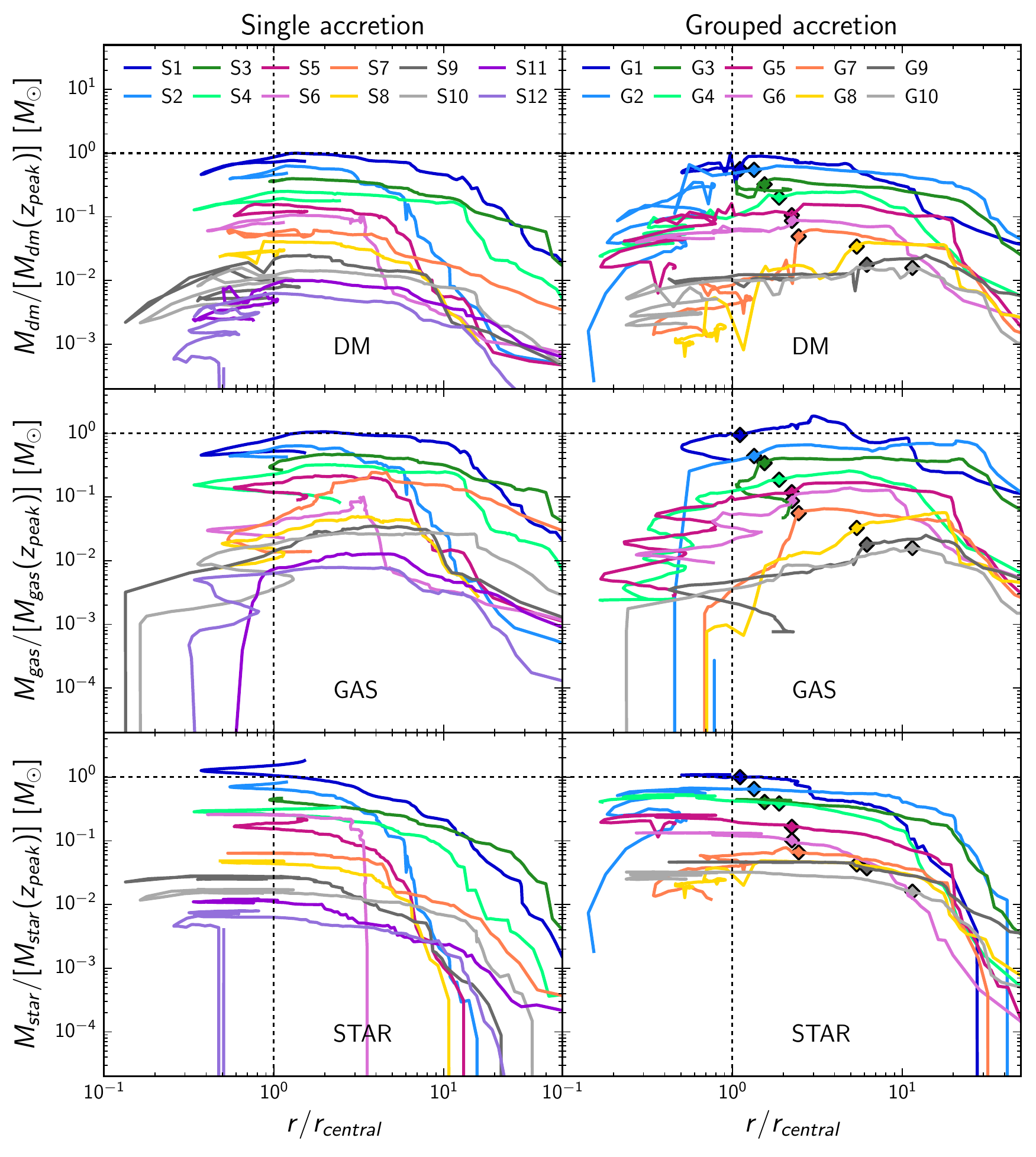}
\caption{Mass in dark matter (top), gas (middle) and stars (bottom), normalized by the corresponding value at $z_{\text{peak}}$, for each of the galaxies as a function of their distance from the main group's centre, normalized by the group's virial radius at the time. To see each trajectory clearly, we have applied a 0.2 dex offset between consecutive galaxies from S1-S12 and G1-G10. The vertical black dashed line indicates the group's virial radius. For the grouped galaxies, we also indicate the first time they become part of any group with diamond symbols.}	\label{fig:massTrajPeak}
\end{figure*}

\begin{figure}
\includegraphics[width=\linewidth]{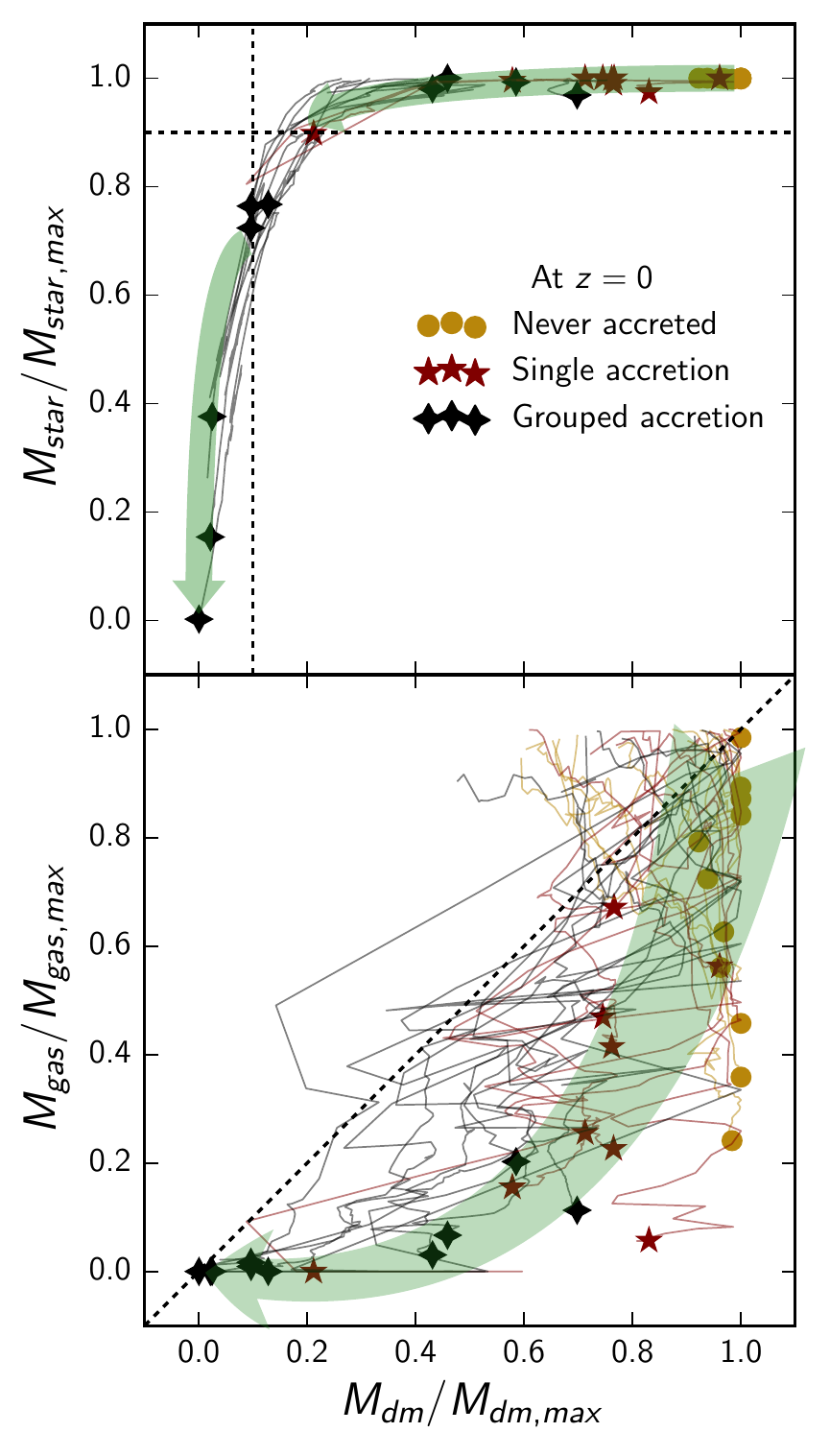}
\caption{Stellar (\emph{top}) and gas (\emph{bottom}) mass as a fraction of their maximum values for each galaxy, versus dark matter mass as a fraction of maximum dark matter mass. Note that the maximum stellar (gas) mass is not the same as the stellar (gas) mass at peak total mass as the maximum in dark matter mass, gas mass and stellar mass can occur at different times for a single galaxy. The coloured points show the values at $z=0$. The faint lines show the trajectories of each galaxy in these planes, starting from the time at which each galaxy reaches its maximum stellar and gas mass respectively. The line colours correspond to the galaxy sample. We have added faint green arrows to indicate a general evolutionary track for all galaxies. A large fraction of the unaccreted galaxies reached their maximum stellar mass very recently, resulting in very short tracks in these plots. The vertical and horizontal dashed lines in the top panel show 10\% remaining dark matter mass and 90\% remaining stellar mass respectively; the dashed line in the bottom panel delineates equal amounts of gas and dark matter retained.}	\label{fig:massFracLoss}
\end{figure}

\begin{figure*}
\includegraphics[width=\linewidth]{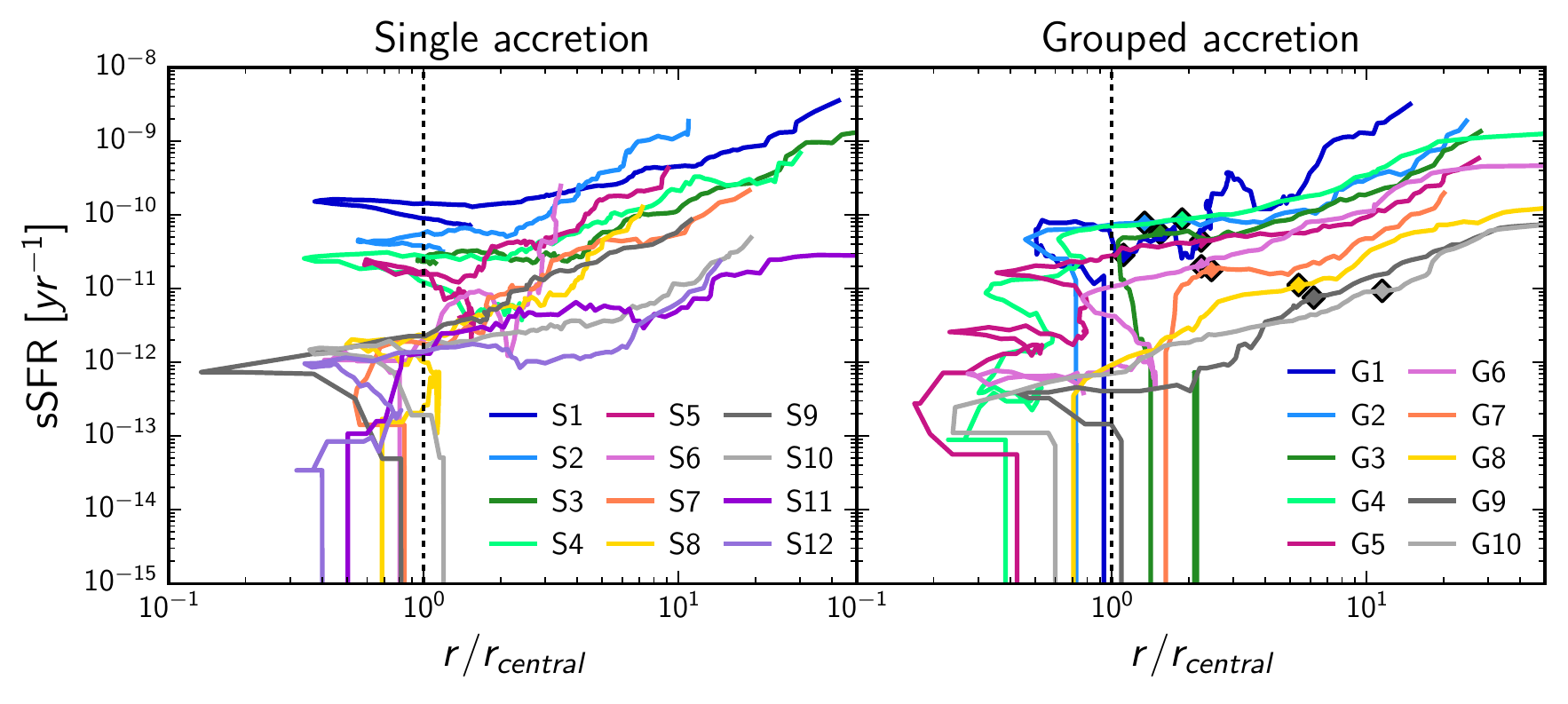}
\caption{sSFR of each of the galaxies as a function of their distance from the main group's centre, normalized by the group's virial radius. The colours and symbols are the same as in Fig. \ref{fig:massTrajPeak}. To see each trajectory clearly, we have applied a 0.2 dex offset between consecutive galaxies from S1-S12 and G1-G10. The data have been smoothed using a moving average of width 10 for clarity.}	\label{fig:ssfrTraj} 
\end{figure*}

\begin{figure*}
\includegraphics[width=\linewidth]{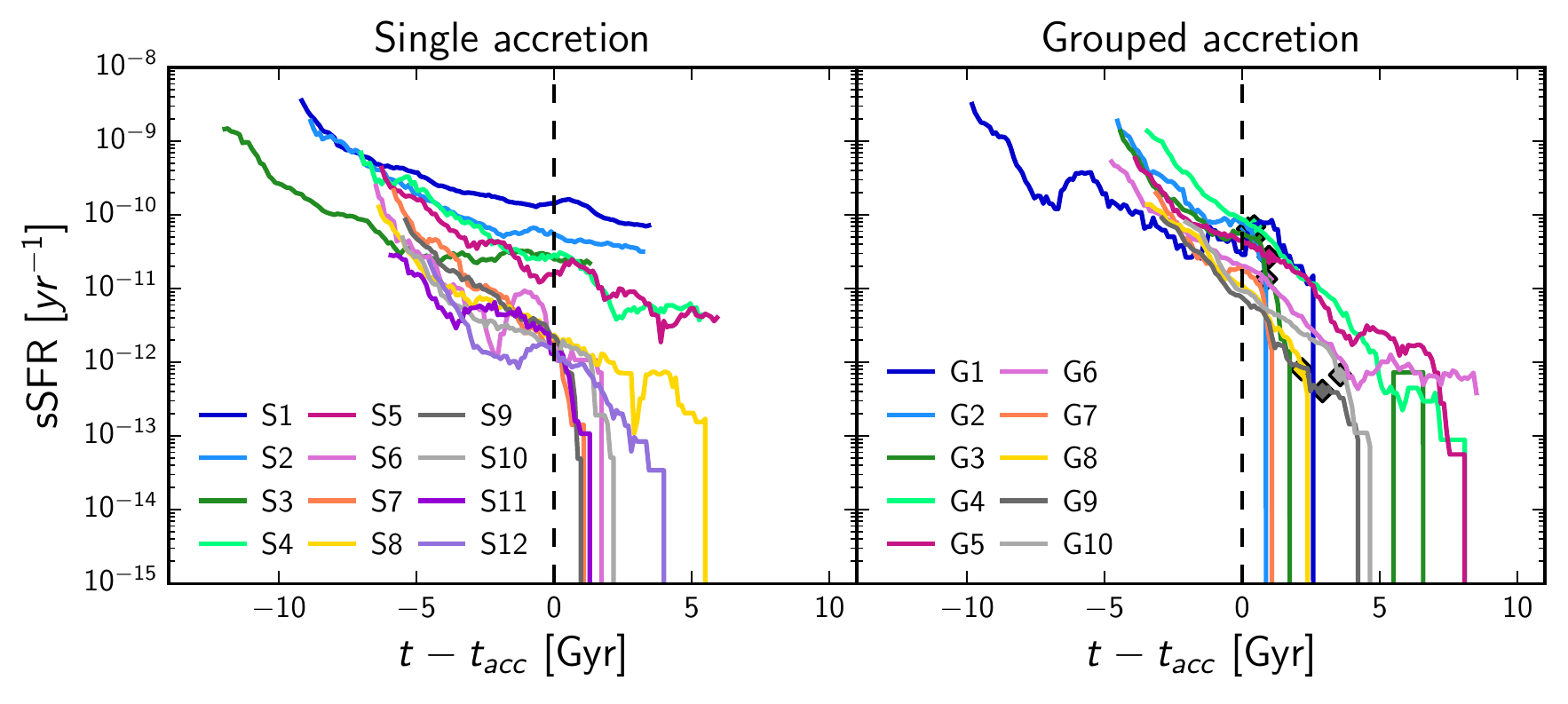}
\caption{sSFR of each galaxy as a function of time since accretion onto any halo. The colours are the same as in Fig. \ref{fig:massTrajPeak}. The diamond symbols indicate when the grouped galaxies first crossed within the \emph{main} group's virial radius. To see each trajectory clearly, we have applied a 0.2 dex offset between consecutive galaxies from S1-S12 and G1-G10. The data have been smoothed using a moving average of width 10.}	\label{fig:ssfrVsTime}
\end{figure*}

The efficiency of the processes that can affect the single and grouped galaxies can vary significantly with location within the main group. Therefore, following the trajectories of these galaxies can shed light on how their properties are affected by their proximity to the group. 

In Fig. \ref{fig:massTrajPeak}, we show the evolution of the galaxies' mass in dark matter (top panels), gas (middle panels) and stars (bottom panels) as a function of their distance from the main group centre, with the masses normalized by their respective values at $z_{\text{peak}}$. The top left panel shows clearly how the single galaxies' dark matter masses remain nearly constant after peak until they cross within $r_{\text{vir}}$ of the group. Grouped galaxies, shown in the top right panel, are accreted by external groups approximately at the same time they reach their peak mass, and begin mass loss soon after this. The degree of total mass loss since peak is higher for grouped galaxies than for single galaxies on average. The middle panels of Fig. \ref{fig:massTrajPeak} show qualitatively the same trends for gas mass and also confirm that the peak in total mass (which is dominated by dark matter) occurs at roughly the same time as the peak in gas mass for most of the galaxies. The bottom panels of Fig. \ref{fig:massTrajPeak} show that both galaxy samples continue growing in stellar mass for some time after they reach their peak total mass. Note however that there a few galaxies that show a significant loss in stellar mass -- S12, G2, G7 and G8. These are also galaxies that have lost over 90\% of their dark matter since $z_{\text{peak}}$.

This difference in loss of stars versus dark matter is shown more explicitly in the top panel of Fig. \ref{fig:massFracLoss}, where we show the stellar mass as a fraction of the maximum stellar mass versus the dark matter mass as a fraction of the maximum dark matter mass. Note that the maximum masses are not necessarily the same as the mass at $z_{\text{peak}}$ since the maximum dark matter mass, gas mass and stellar mass can occur at different times for a single galaxy. Individual galaxy trajectories are shown in faint lines, starting from the time when the galaxies reach their maximum stellar mass, thereby showing only the phase of mass decline. The coloured symbols mark the end of these trajectories at $z=0$. The figure shows that galaxies that have lost $<90\%$ of the maximum dark matter mass still retain $>90\%$ of their maximum stellar mass. These results are broadly consistent with work by \citet{Smith16} who showed that galaxies around clusters that lose $\sim80\%$ of their dark matter mass only lose $\sim10\%$ of their stellar mass. Our results are not directly comparable since we do not follow the galaxies' particles from peak to $z=0$ in the same way as \citet{Smith16}.

The bottom panel of Fig. \ref{fig:massFracLoss} similarly shows the gas mass as a fraction of the maximum gas mass versus dark matter mass as a fraction of its maximum. The galaxies lose their gas mass more rapidly than dark matter, although there is significant variation between the galaxies.

We show the trajectories in sSFRs of our galaxy sample in Fig. \ref{fig:ssfrTraj}. The sSFRs for both sets of galaxies gradually decline as they approach the main group. Grouped galaxies do not show a sudden drop in sSFR when they are accreted by external groups. Contrary to what was seen in Fig. \ref{fig:ssfrEvol} however, several of the galaxies do show a sharp decline in sSFR some time after entering the main group. It must be kept in mind that there is no one-to-one correlation between the galaxies' distance from the group and cosmic time and what appears as a rapid decline could be occurring over a larger time interval. We therefore also show the sSFRs as a function of time in Fig. \ref{fig:ssfrVsTime}. The sSFRs for roughly half of the single galaxies and almost of the grouped galaxies show a sharp decline, indicating rapid quenching, after a delay ranging from $\sim(0.5-2.5)$ Gyr after accretion onto the main group. Galaxies S8, S12, G5, G9 and G10 deviate from this behaviour, showing delays of $>4$ Gyr between accretion onto the main group and quenching. Although this is only a small sample of galaxies in one group, these results are broadly consistent with recent proposed models of quenching following a delay of $(2-3)$ Gyr after accretion. \citep[e.g.][]{Wetzel13,Schawinski14,Balogh16,Oman16}. It is important to note that for almost all the grouped galaxies that are quenched, the final quenching event occurs after accretion onto the main group. For roughly half of them, this is also within the $\sim(0.5-2.5)$ Gyr interval after accretion onto their external groups. It is therefore difficult to determine which environment is the dominant cause for quenching. However, the fraction of grouped galaxies that are quenched at $z=0$ is higher than the single galaxies, and on average, the grouped galaxies have significantly lower sSFRs than the single galaxies.

\section{Mass loss mechanisms}	\label{sec:lossMech}

\begin{figure*}
\includegraphics[width=\linewidth]{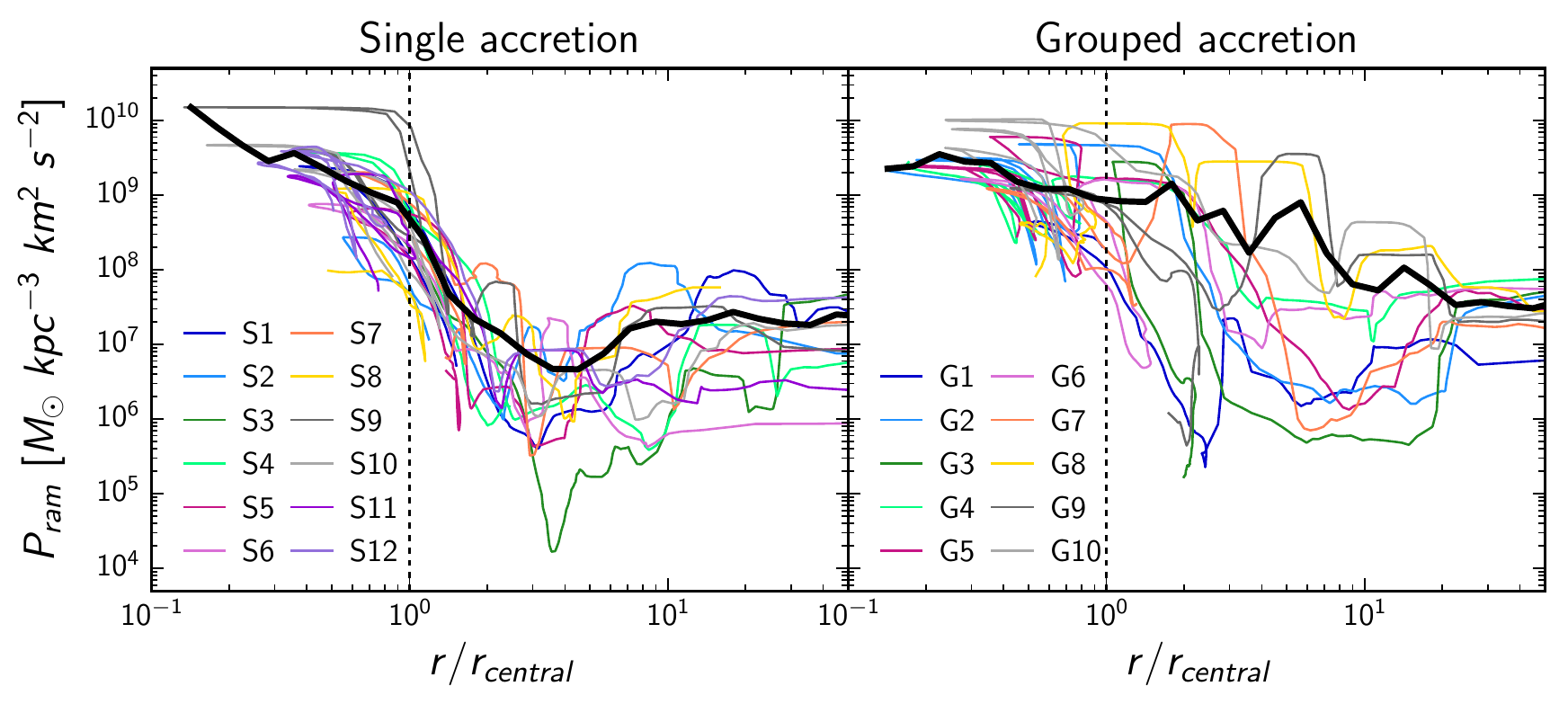}
\caption{Ram pressure acting on each galaxy as a function of distance from the main group. The colours are the same as in Fig. \ref{fig:massTrajPeak}. Faint coloured lines show the trajectories of individual galaxies, while the heavy black lines show the average trajectory for each subsample. The data have been smoothed using a moving average of width 10 for clarity.} \label{fig:ramPress}
\end{figure*}

\begin{figure*}
\includegraphics[width=\linewidth]{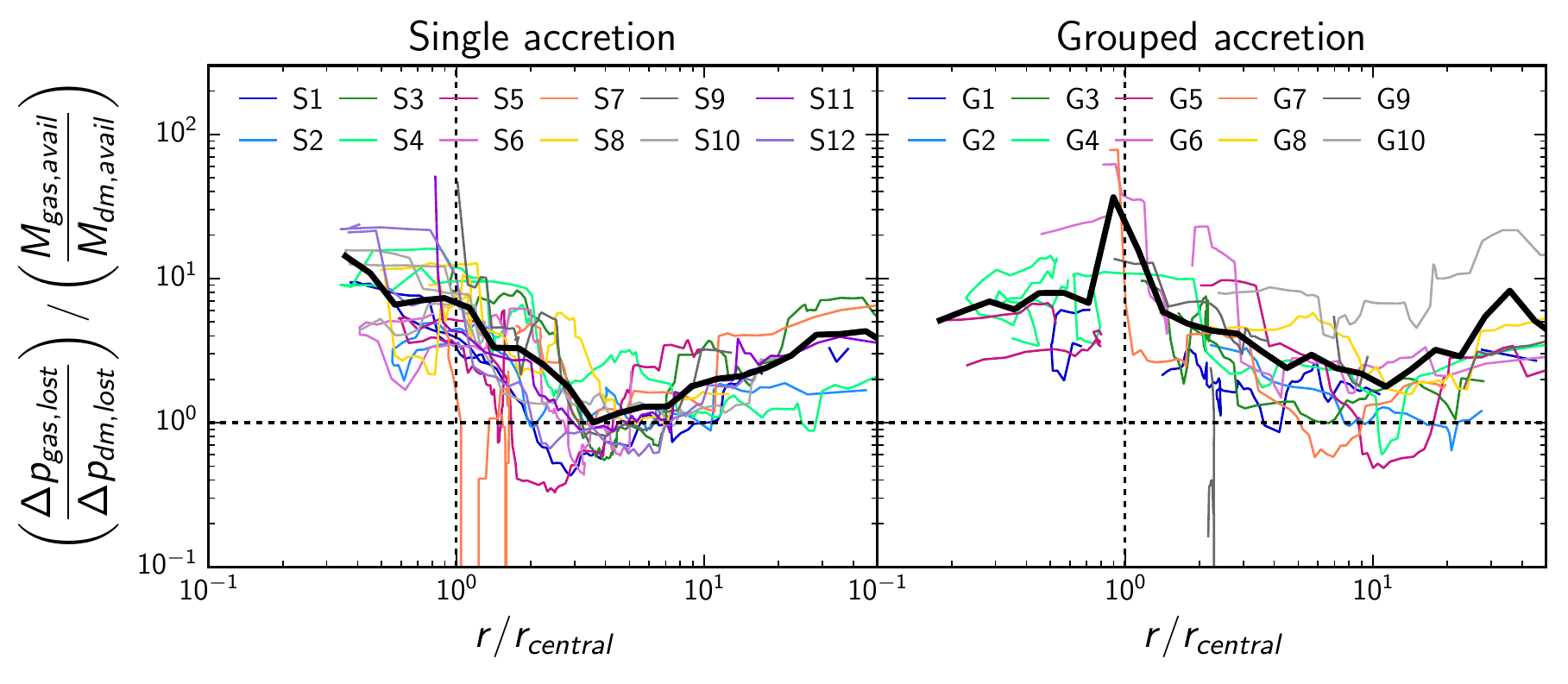}
\caption{Ratio of the change in momentum of gas particles \emph{lost} at each timestep and change in momentum of lost dark particles, normalized by the ratio of mass in gas to dark matter available in the previous snapshot, as a function of distance from the main group. Faint coloured lines show the trajectories of individual galaxies, while the heavy black lines show the average trajectory for each subsample. The data have been smoothed using a moving average of width 10 for clarity.} \label{fig:dmVsGas}
\end{figure*}

As mentioned in the introduction, several environmental mechanisms can result in the evolution of properties we see for both single and grouped galaxies. In order to disentangle which of these mechanisms is dominant in this system, we examine the differences between the dark matter and gas components of our galaxies. While the loss of dark matter can largely be attributed to tidal truncation, the loss of gas mass could be the result of a combination of processes. One process that can preferentially remove gas from a galaxy without removing dark matter is ram pressure stripping. Fig. \ref{fig:ramPress} shows the ram pressure experienced by the galaxies as they approach the group. We calculate ram pressure as
\begin{equation}
P_{\text{ram}} = \rho \bar{v}^{2}
\end{equation}
where $\rho$ is the gas density within a 10 kpc wide shell outside the tidal radius of the galaxy and $\bar{v}$ is the average velocity of  the gas particles within this shell, relative to the galaxy's own velocity. This allows us to directly sample the properties of the gas in the immediate vicinity of the galaxy.

For the single galaxies, ram pressure increases once they are within $(1-3)\,r_{\text{vir}}$ of the group. The grouped galaxies experience similar ram pressure as the single galaxies within $r_{\text{vir}}$; outside $r_{\text{vir}}$, however, the ram pressure out to $\sim(5-7)\,r_{\text{vir}}$ is higher than that experienced by the single galaxies, which is also when the grouped galaxies are within their external groups.

The amount of ram pressure experienced by the galaxies is only one side of the equation; whether or not it is strong enough to overcome the gravitational force exerted by the galaxy is difficult to determine directly. The ideal method is to compare the ram pressure to the rate of change of momentum of the galaxies' gas particles; however this method requires several assumptions to be made, such as the radius within the galaxy where ram pressure is effective. Instead, in Fig. \ref{fig:dmVsGas}, we show the ratio of change in momentum of \emph{lost} gas particles to that of the lost dark matter particles for each galaxy, normalized by the ratio of gas mass to dark matter mass available in the previous snapshot. Lost particles are those particles that were assigned to the galaxy in the previous snapshot, but not in the current snapshot. Note that at all times, at least $80\%$ and often $90\%$ of the gas content of all galaxies in our sample is in cold gas -- the results for cold gas are similar to the results in Fig. \ref{fig:dmVsGas} and the loss of hot gas mass is a small contribution to the overall loss of gas mass. Therefore, we do not show the two components separately. 

Fig. \ref{fig:ramPress} should be interpreted with caution -- while the loss of momentum through dark matter particles probes the effect of the gravitational forces, the additional loss of momentum through gas particles is due to a variety of hydrodynamical effects which include ram pressure. It is clear from Fig. \ref{fig:ramPress} that both samples of galaxies lose relatively more momentum through the loss of gas particles. Although we cannot determine the precise amount of gas lost specifically through ram pressure stripping, the similarity between the profiles for the ratio of momentum of lost gas and dark matter particles for the galaxies and the ram pressure they experience shown in Fig. \ref{fig:ramPress} indicates that ram pressure stripping is a plausible scenario to explain these results.

Combining these results with those from Sections \ref{sec:evol} and \ref{sec:traj}, we can infer the following:
\begin{itemize}
\item Tidal truncation plays the biggest role in the removal of mass from these galaxies. The stellar mass of the galaxies, which is tightly bound near the centres of the galaxies, is not significantly stripped until the galaxies lose a large portion of their dark matter mass. However, both the dark matter and gas components are affected significantly, as seen in Figs. \ref{fig:massEvolNorm} and \ref{fig:massTrajPeak}.
\item Several of the galaxies preferentially lose more gas mass than dark matter mass, as seen in Fig. \ref{fig:massTrajPeak}. While it is difficult to precisely determine how much of the gas lost is due to ram pressure stripping, we find that when the ram pressure experienced by the galaxies rises, they lose relatively more momentum through lost gas particles compared to dark matter particles.
\end{itemize}

\section{Implications for galaxy evolution}	\label{sec:implications}

\begin{figure}
\includegraphics[width=\linewidth]{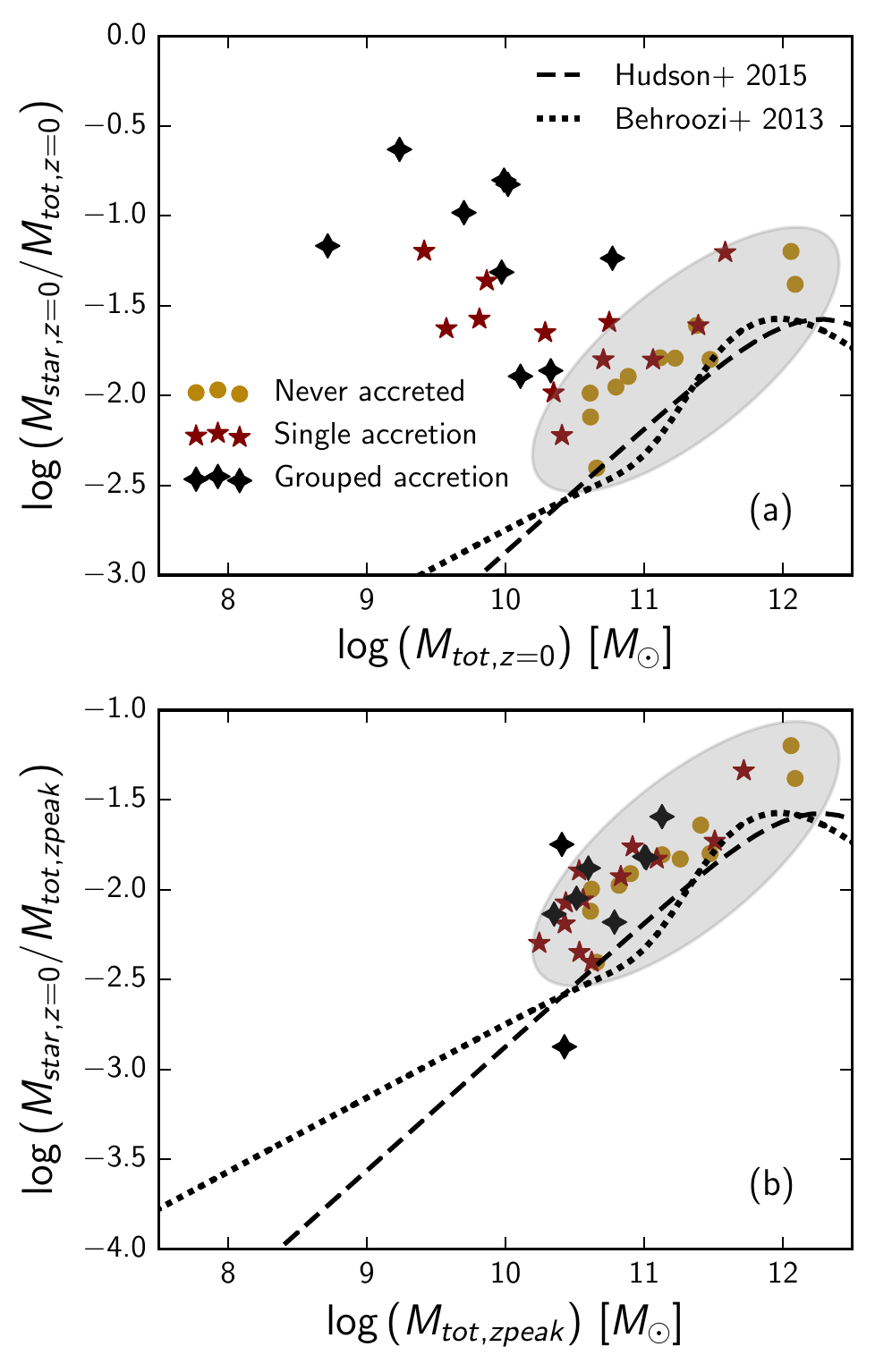}
\caption{Stellar fraction w.r.t. total mass at $z=0$ (top) and at $z_{\text{peak}}$ (bottom) for galaxies in all three categories. The grey ellipse has been added in the top panel to approximately show the region occupied by the unaccreted galaxies. The elliptical region in the bottom panel is identical to the one above. The figures show explicitly that the current stellar mass of the galaxies is better correlated with their peak total mass rather than present day mass.} \label{fig:smhmRelation}
\end{figure}

\begin{figure}
\centering
\includegraphics[width=\linewidth]{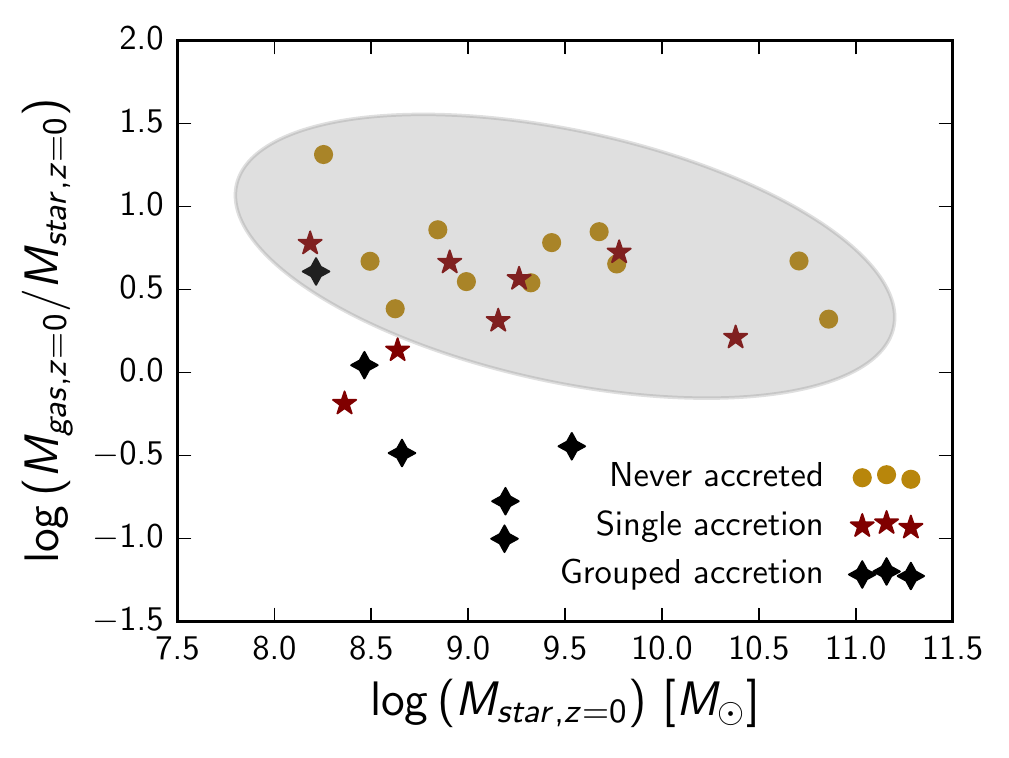}
\caption{Gas fraction as a function of stellar mass at $z=0$ for galaxies in all three categories. The  grey ellipse has been added to approximately show the region occupied by the unaccreted galaxies.} \label{fig:gasFraction}
\end{figure}

Finally, we show the consequences of these environmental processes on the final properties of our galaxy sample. In Fig. \ref{fig:smhmRelation}, we show the stellar fraction $M_{\text{star}}/M_{\text{tot}}$ versus $M_{\text{tot}}$ for our galaxies, where $M_{\text{tot}}$ is the total mass at $z=0$ (top panel) and at $z_{\text{peak}}$ (bottom panel), while $M_{\text{star}}$ is always the stellar mass at $z=0$. In Fig. \ref{fig:smhmRelation}(a), unaccreted galaxies show a strong correlation between stellar mass fraction and total mass at $z=0$ as expected. However, there is a clear difference between the unaccreted and grouped galaxies. As can be seen in Fig. \ref{fig:massTrajPeak}, as galaxies are accreted onto a group, they lose dark matter mass, but not stellar mass. This results in grouped galaxies of the same total mass having higher stellar fractions on average compared to unaccreted galaxies by as much as two orders of magnitude. The single galaxies occupy an intermediate region in the SMHM plane between the unaccreted and grouped galaxies. Thus the grouped galaxies occupy a distinct region on this diagram compared to unaccreted galaxies. The results of Fig. \ref{fig:smhmRelation}(a) indicate that preprocessing has a significant impact on the present day stellar mass-halo mass (SMHM) relations. The differences between unaccreted, single and grouped galaxies represent a significant source of scatter in the stellar mass-halo mass relation for any sample that includes group/cluster galaxies. It should be noted that these results do depend on the fact that we define galaxies by their tidal radii. The dark matter content of the galaxies is disproportionately affected by this truncation compared to the stellar component. However, as can be seen in Fig. \ref{fig:massTrajPeak}, galaxy tidal radii and mass grow again after pericentric passages, allowing temporarily unbound material to become bound to the galaxies again. Therefore, the use of a tidal radius only has a significant effect close to the centre of the group. We have confirmed that these trends still hold even when using the virial radii provided by \textsc{rockstar} to define the extent of the galaxies.

We also show two SMHM relations derived using observational data in Fig. \ref{fig:smhmRelation} for comparison. The dashed line shows the relation from \citet{Hudson15}, derived using weak lensing data from CFHTLenS. We show their default model fit for their entire sample of galaxies. The dotted line shows the relation from \citet{Behroozi13c}, which was modelled to match several observational constraints on stellar mass functions, sSFRs of galaxies and the cosmic SFR from the literature. We calculate both relations at $z=0$. Fig. \ref{fig:smhmRelation} shows that our sample of unaccreted galaxies is in good agreement with both observational results, while most of the single galaxies and all grouped galaxies deviate significantly from the observed relations.

Previous studies have suggested a strong correlation between the stellar mass and peak mass of galaxies \citep[e.g.][]{Penarrubia08,Smith16}. As we have found in this study, most of the single and grouped galaxies reach their peak total mass around the time they first became part of any group. Before accretion, their stellar fractions evolve in the same way as unaccreted galaxies. In Fig. \ref{fig:smhmRelation}(b), we find that all three categories of galaxies follow a single trend in stellar fraction versus total mass at $z_{\text{peak}}$. In fact, this trend is nearly identical to the trend seen for unaccreted galaxies at $z=0$, which indicates that there is indeed a strong correlation between current stellar mass and halo mass at $z_\text{{peak}}$. The one outlier to this trend is galaxy G2, which has lost nearly all of its stellar mass along with its dark matter and gas. Overall, we find that the present day stellar mass of the galaxies in our sample is much more strongly correlated with their peak total mass than with their present day total mass. This result is in agreement with the now standard methodology of Subhalo Abundance Matching (SHAM), whereby stellar masses are assigned to subhaloes in dark matter simulations based on their mass at accretion (which we find to roughly coincide with peak mass) or a related property such as the maximum circular velocity at accretion or at peak mass \cite[e.g. see][]{Vale06,Conroy06,Berrier06,Behroozi10,Chaves16}.

We also investigated the relation between gas fraction and total mass. The gas fraction is well correlated with $M_{\text{tot}}$ at both $z_{\text{peak}}$ and $z=0$ and there was no obvious separation between the three samples of galaxies. Observationally, the total mass of galaxies is often difficult to measure, making it challenging to compare our results to observations, whereas the stellar and gas mass are more easily obtained. Therefore, in Fig. \ref{fig:gasFraction}, we show the gas fraction w.r.t. stellar mass, $M_{\text{gas}}/M_{\text{star}}$, as a function of $M_{\text{star}}$ at $z=0$. Here again, the grouped galaxies occupy a distinct region in this space compared to unaccreted galaxies, which shows that the gas fraction is significantly affected by environment.

\section{Summary}	\label{sec:summ}

We use a zoom-in hydrodynamical simulation of a galaxy group out to $3\,r_{\text{vir}}$ to study the mass loss histories of its member galaxies and the degree to which they have been preprocessed.

\begin{itemize}
\item Both singly-accreted and group-accreted galaxies lose considerable amounts of mass in dark matter and gas due to their accretion onto a group. The mass loss is primarily due to tidal truncation which does not significantly affect the stellar content of most of the galaxies.
\item Single and grouped galaxies both reach their peak total mass before becoming part of the main group. Single galaxies have a nearly constant mass in dark matter and gas after this peak, which occurs at $\sim(1-3)\,r_{\text{vir}}$ from the group, until they begin losing mass just outside $r_{\text{vir}}$. Grouped galaxies begin losing mass well outside $r_{\text{vir}}$ of the main group due to the external groups they are in. This preprocessing results in more total mass loss from peak to $z=0$ than single galaxies.
\item Both single and grouped galaxies have gradually declining sSFRs as they approach the main group -- we find evidence for rapid quenching for most galaxies from both samples following a delay of  $\sim(0.5-2.5)$ Gyr after accretion onto \emph{any} group. In the case of the grouped galaxies, it is unclear which environment, their external groups or the main group, is primarily responsible for quenching.
\item In most cases, galaxies lose relatively more mass in gas than in dark matter and this cannot be accounted for with star formation alone. We find that when the galaxies experience higher amounts of ram pressure, they lose relatively more momentum through lost gas particles than dark matter particles, suggesting that ram pressure may play a significant secondary role in gas removal.
\item The cumulative effect of this mass evolution is that at the final redshift, unaccreted and grouped galaxies occupy distinct regions on a SMHM relation. In fact, preprocessed galaxies are likely to be an important source of scatter in any SMHM that does not exclude group and cluster galaxies. At peak mass, however, all three categories adhere to a single SMHM relation and therefore, peak masses can be reliably used to assign stellar masses to haloes in dark matter simulations, regardless of their accretion histories.
\end{itemize}

This study examines the mass loss of galaxies in a group environment and the role preprocessing plays in mass loss and star formation. We intend to expand this study to more groups, in order to build a statistical understanding of galaxy evolution in groups. Our future work will focus not only on expanding our sample, but also extending this analysis to observable properties that can be directly compared to observations of galaxies in groups.

\section*{Acknowledgements}
We thank the anonymous referee for their comments and insights, which greatly helped in improving the manuscript. We thank the National Science and Engineering Research Council of Canada for their funding. Computations were performed on the \textsc{gpc} supercomputer at the SciNet HPC Consortium \citep{Loken10}. SciNet is funded by the following: the Canada Foundation for Innovation under the auspices of Compute Canada; the Government of Ontario; Ontario Research Fund -- Research Excellence; and the University of Toronto. This work was made possible by the facilities of the Shared Hierarchical Academic Research Computing Network (SHARCNET:www.sharcnet.ca) and Compute Canada. BWK gratefully acknowledges funding from the European Research Council (ERC) under the European Union's Horizon 2020 research and innovation programme via the ERC Starting Grant MUSTANG (grant agreement number 714907, PI Kruijssen), and from the Alexander von Humboldt Stiftung via a Postdoctoral Research Fellowship. Additionally, this work made use of several open-source software packages such as \textsc{Pynbody} \citep{pynbody}, \textsc{Numpy} \citep{numpy} and \textsc{Matplotlib} \citep{matplotlib}.

\bibliographystyle{mnras}
\bibliography{JoshiParkerWadsleyKeller2018}


\bsp	
\label{lastpage}
\end{document}

%% file: figures/distinct_properties.tex
N1 & 290.2 & $1.2 \times 10^{12}$ & $9.4 \times 10^{11}$ & $2.4 \times 10^{11}$ & $5.1 \times 10^{10}$ & 1.8 & - & - \\
N2 & 291.1 & $1.1 \times 10^{12}$ & $9.2 \times 10^{11}$ & $1.5 \times 10^{11}$ & $7.3 \times 10^{10}$ & 2.4 & - & - \\
N3 & 181.7 & $3.0 \times 10^{11}$ & $2.6 \times 10^{11}$ & $3.3 \times 10^{10}$ & $4.7 \times 10^{9}$ & 2.1 & - & - \\
N4 & 145.4 & $2.4 \times 10^{11}$ & $2.1 \times 10^{11}$ & $2.6 \times 10^{10}$ & $5.8 \times 10^{9}$ & 2.1 & - & - \\
N5 & 135.0 & $1.7 \times 10^{11}$ & $1.5 \times 10^{11}$ & $1.6 \times 10^{10}$ & $2.7 \times 10^{9}$ & 1.1 & - & - \\
N6 & 128.4 & $1.3 \times 10^{11}$ & $1.2 \times 10^{11}$ & $7.3 \times 10^{9}$ & $2.1 \times 10^{9}$ & 1.9 & - & - \\
N7 & 113.9 & $7.7 \times 10^{10}$ & $7.2 \times 10^{10}$ & $3.5 \times 10^{9}$ & $9.8 \times 10^{8}$ & 1.5 & - & - \\
N8 & 108.6 & $6.3 \times 10^{10}$ & $5.7 \times 10^{10}$ & $5.0 \times 10^{9}$ & $7.0 \times 10^{8}$ & 2.4 & - & - \\
N9 & 99.3 & $4.5 \times 10^{10}$ & $4.2 \times 10^{10}$ & $3.7 \times 10^{9}$ & $1.8 \times 10^{8}$ & 2.8 & - & - \\
N10 & 98.7 & $4.1 \times 10^{10}$ & $3.9 \times 10^{10}$ & $1.5 \times 10^{9}$ & $3.1 \times 10^{8}$ & 2.3 & - & - \\
N11 & 87.1 & $4.1 \times 10^{10}$ & $3.9 \times 10^{10}$ & $1.0 \times 10^{9}$ & $4.2 \times 10^{8}$ & 2.4 & - & - \\

%% file: figures/single_properties.tex
S1 & 190.6 & $3.8 \times 10^{11}$ & $3.2 \times 10^{11}$ & $3.9 \times 10^{10}$ & $2.4 \times 10^{10}$ & 0.4 & 0.29 & - \\
S2 & 169.9 & $2.5 \times 10^{11}$ & $2.1 \times 10^{11}$ & $3.2 \times 10^{10}$ & $6.0 \times 10^{9}$ & 0.6 & 0.27 & - \\
S3 & 126.3 & $1.2 \times 10^{11}$ & $1.1 \times 10^{11}$ & $6.7 \times 10^{9}$ & $1.8 \times 10^{9}$ & 0.9 & 0.10 & - \\
S4 & 103.2 & $5.6 \times 10^{10}$ & $5.2 \times 10^{10}$ & $2.9 \times 10^{9}$ & $1.4 \times 10^{9}$ & 0.3 & 0.53 & - \\
S5 & 97.3 & $5.1 \times 10^{10}$ & $4.6 \times 10^{10}$ & $3.7 \times 10^{9}$ & $8.0 \times 10^{8}$ & 0.6 & 0.61 & - \\
S6 & 83.9 & $2.6 \times 10^{10}$ & $2.4 \times 10^{10}$ & $9.1 \times 10^{8}$ & $1.5 \times 10^{8}$ & 0.4 & 0.19 & - \\
S7 & 75.2 & $2.2 \times 10^{10}$ & $2.2 \times 10^{10}$ & $1.5 \times 10^{8}$ & $2.3 \times 10^{8}$ & 0.5 & 0.75 & - \\
S8 & 39.1 & $1.9 \times 10^{10}$ & $1.8 \times 10^{10}$ & $5.9 \times 10^{8}$ & $4.3 \times 10^{8}$ & 0.5 & 0.61 & - \\
S9 & 44.8 & $7.3 \times 10^{9}$ & $7.0 \times 10^{9}$ & 0.0 & $3.2 \times 10^{8}$ & 0.1 & 0.79 & - \\
S10 & 41.9 & $6.5 \times 10^{9}$ & $6.3 \times 10^{9}$ & 0.0 & $1.7 \times 10^{8}$ & 0.2 & 0.77 & - \\
S11 & 30.3 & $3.7 \times 10^{9}$ & $3.7 \times 10^{9}$ & 0.0 & $8.8 \times 10^{7}$ & 0.3 & 0.86 & - \\
S12 & 23.9 & $2.6 \times 10^{9}$ & $2.4 \times 10^{9}$ & 0.0 & $1.7 \times 10^{8}$ & 0.3 & 0.96 & - \\

%% file: figures/grouped_properties.tex
G1 & 56.7 & $1.3 \times 10^{10}$ & $1.2 \times 10^{10}$ & $6.7 \times 10^{8}$ & $1.6 \times 10^{8}$ & 0.5 & 0.19 & 0.22 \\
G2 & 2.4 & $8.2 \times 10^{6}$ & $7.7 \times 10^{6}$ & 0.0 & $4.4 \times 10^{5}$ & 0.1 & 0.93 & 0.98 \\
G3 & 74.4 & $2.1 \times 10^{10}$ & $2.1 \times 10^{10}$ & $3.2 \times 10^{8}$ & $2.9 \times 10^{8}$ & 1.1 & 0.93 & 1.04 \\
G4 & 14.9 & $9.8 \times 10^{9}$ & $8.1 \times 10^{9}$ & $1.5 \times 10^{8}$ & $1.5 \times 10^{9}$ & 0.2 & 1.44 & 1.62 \\
G5 & 31.4 & $1.0 \times 10^{10}$ & $8.6 \times 10^{9}$ & $2.6 \times 10^{8}$ & $1.6 \times 10^{9}$ & 0.2 & 0.93 & 1.18 \\
G6 & 73.3 & $5.9 \times 10^{10}$ & $5.4 \times 10^{10}$ & $1.2 \times 10^{9}$ & $3.4 \times 10^{9}$ & 0.3 & 0.93 & 1.15 \\
G7 & 19.2 & $5.2 \times 10^{8}$ & $4.9 \times 10^{8}$ & 0.0 & $3.6 \times 10^{7}$ & 0.3 & 0.82 & 1.29 \\
G8 & 18.8 & $1.7 \times 10^{9}$ & $1.3 \times 10^{9}$ & 0.0 & $4.0 \times 10^{8}$ & 0.5 & 0.93 & 1.62 \\
G9 & 55.8 & $9.4 \times 10^{9}$ & $8.8 \times 10^{9}$ & $1.5 \times 10^{8}$ & $4.6 \times 10^{8}$ & 0.4 & 0.88 & 1.90 \\
G10 & 26.5 & $5.0 \times 10^{9}$ & $4.5 \times 10^{9}$ & 0.0 & $5.2 \times 10^{8}$ & 0.2 & 0.93 & 2.52 \\

%% file: figures/distinct_properties_peak.tex
N1 & 0.01 & 1.8 & 290.2 & $1.2 \times 10^{12}$ & $9.4 \times 10^{11}$ & $2.4 \times 10^{11}$ & $5.1 \times 10^{10}$ \\
N2 & 0.02 & 2.5 & 290.2 & $1.1 \times 10^{12}$ & $9.2 \times 10^{11}$ & $1.5 \times 10^{11}$ & $7.1 \times 10^{10}$ \\
N3 & 0.01 & 2.1 & 181.7 & $3.0 \times 10^{11}$ & $2.6 \times 10^{11}$ & $3.3 \times 10^{10}$ & $4.7 \times 10^{9}$ \\
N4 & 0.25 & 2.1 & 149.2 & $2.6 \times 10^{11}$ & $2.2 \times 10^{11}$ & $3.1 \times 10^{10}$ & $4.8 \times 10^{9}$ \\
N5 & 0.07 & 1.4 & 146.8 & $1.8 \times 10^{11}$ & $1.6 \times 10^{11}$ & $1.9 \times 10^{10}$ & $2.6 \times 10^{9}$ \\
N6 & 0.02 & 1.9 & 139.7 & $1.3 \times 10^{11}$ & $1.3 \times 10^{11}$ & $7.3 \times 10^{9}$ & $2.1 \times 10^{9}$ \\
N7 & 0.50 & 1.5 & 93.9 & $8.0 \times 10^{10}$ & $7.2 \times 10^{10}$ & $7.0 \times 10^{9}$ & $6.8 \times 10^{8}$ \\
N8 & 0.19 & 3.1 & 101.5 & $6.6 \times 10^{10}$ & $5.9 \times 10^{10}$ & $6.6 \times 10^{9}$ & $6.0 \times 10^{8}$ \\
N9 & 0.01 & 2.8 & 99.3 & $4.5 \times 10^{10}$ & $4.2 \times 10^{10}$ & $3.7 \times 10^{9}$ & $1.8 \times 10^{8}$ \\
N10 & 0.01 & 2.3 & 98.7 & $4.1 \times 10^{10}$ & $3.9 \times 10^{10}$ & $1.5 \times 10^{9}$ & $3.1 \times 10^{8}$ \\
N11 & 0.13 & 3.0 & 86.8 & $4.2 \times 10^{10}$ & $4.0 \times 10^{10}$ & $1.3 \times 10^{9}$ & $4.1 \times 10^{8}$ \\

%% file: figures/single_properties_peak.tex
S1 & 0.35 & 1.4 & 177.9 & $5.2 \times 10^{11}$ & $4.3 \times 10^{11}$ & $7.9 \times 10^{10}$ & $1.3 \times 10^{10}$ \\
S2 & 0.31 & 1.2 & 155.9 & $3.2 \times 10^{11}$ & $2.7 \times 10^{11}$ & $4.7 \times 10^{10}$ & $4.5 \times 10^{9}$ \\
S3 & 0.17 & 1.3 & 125.2 & $1.2 \times 10^{11}$ & $1.1 \times 10^{11}$ & $1.0 \times 10^{10}$ & $1.6 \times 10^{9}$ \\
S4 & 0.57 & 1.2 & 88.8 & $8.2 \times 10^{10}$ & $7.2 \times 10^{10}$ & $8.9 \times 10^{9}$ & $1.1 \times 10^{9}$ \\
S5 & 0.59 & 0.8 & 78.4 & $6.8 \times 10^{10}$ & $6.1 \times 10^{10}$ & $6.5 \times 10^{9}$ & $5.8 \times 10^{8}$ \\
S6 & 1.25 & 3.4 & 51.6 & $3.4 \times 10^{10}$ & $3.0 \times 10^{10}$ & $4.0 \times 10^{9}$ & $5.9 \times 10^{7}$ \\
S7 & 0.86 & 1.3 & 54.2 & $2.7 \times 10^{10}$ & $2.6 \times 10^{10}$ & $6.8 \times 10^{8}$ & $2.3 \times 10^{8}$ \\
S8 & 0.84 & 1.7 & 59.7 & $3.4 \times 10^{10}$ & $3.1 \times 10^{10}$ & $3.0 \times 10^{9}$ & $3.6 \times 10^{8}$ \\
S9 & 1.01 & 1.6 & 57.8 & $3.6 \times 10^{10}$ & $3.3 \times 10^{10}$ & $2.8 \times 10^{9}$ & $3.2 \times 10^{8}$ \\
S10 & 0.71 & 0.6 & 55.7 & $2.7 \times 10^{10}$ & $2.5 \times 10^{10}$ & $1.5 \times 10^{9}$ & $1.8 \times 10^{8}$ \\
S11 & 1.01 & 1.7 & 43.5 & $1.8 \times 10^{10}$ & $1.5 \times 10^{10}$ & $2.2 \times 10^{9}$ & $8.4 \times 10^{7}$ \\
S12 & 0.98 & 1.0 & 58.2 & $4.2 \times 10^{10}$ & $3.7 \times 10^{10}$ & $4.4 \times 10^{9}$ & $2.6 \times 10^{8}$ \\

%% file: figures/grouped_properties_peak.tex
G1 & 0.21 & 1.0 & 66.5 & $2.2 \times 10^{10}$ & $2.0 \times 10^{10}$ & $1.8 \times 10^{9}$ & $1.5 \times 10^{8}$ \\
G2 & 1.06 & 1.8 & 44.8 & $2.0 \times 10^{10}$ & $1.8 \times 10^{10}$ & $1.7 \times 10^{9}$ & $1.5 \times 10^{8}$ \\
G3 & 1.12 & 1.8 & 51.7 & $3.2 \times 10^{10}$ & $2.9 \times 10^{10}$ & $2.7 \times 10^{9}$ & $2.7 \times 10^{8}$ \\
G4 & 2.42 & 5.3 & 51.2 & $1.0 \times 10^{11}$ & $8.4 \times 10^{10}$ & $1.6 \times 10^{10}$ & $9.6 \times 10^{8}$ \\
G5 & 1.29 & 2.6 & 69.2 & $1.0 \times 10^{11}$ & $8.7 \times 10^{10}$ & $1.5 \times 10^{10}$ & $1.3 \times 10^{9}$ \\
G6 & 1.22 & 2.3 & 73.6 & $1.3 \times 10^{11}$ & $1.2 \times 10^{11}$ & $1.3 \times 10^{10}$ & $2.6 \times 10^{9}$ \\
G7 & 1.44 & 3.2 & 41.9 & $2.7 \times 10^{10}$ & $2.3 \times 10^{10}$ & $3.7 \times 10^{9}$ & $1.9 \times 10^{8}$ \\
G8 & 1.84 & 6.4 & 45.9 & $6.1 \times 10^{10}$ & $5.3 \times 10^{10}$ & $6.7 \times 10^{9}$ & $8.8 \times 10^{8}$ \\
G9 & 2.98 & 13.8 & 30.6 & $2.5 \times 10^{10}$ & $2.0 \times 10^{10}$ & $4.9 \times 10^{9}$ & $2.5 \times 10^{8}$ \\
G10 & 2.62 & 11.5 & 34.2 & $4.0 \times 10^{10}$ & $3.4 \times 10^{10}$ & $5.1 \times 10^{9}$ & $3.3 \times 10^{8}$ \\